\documentclass[aps,pre,notitlepage,floatfix]{revtex4-1}
\usepackage{amsmath}
\usepackage{graphicx}
\usepackage{color}

\begin{document}

\title{On the role of metastable intermediate states in the homogeneous nucleation of solids from solution\footnote{To appear in Advances in Chemical Physics}}
\author{James F. Lutsko}
\affiliation{Center for Nonlinear Phenomena and Complex Systems CP 231, Universit\'{e} Libre de Bruxelles, Blvd. du
Triomphe, 1050 Brussels, Belgium}
\email{jlutsko@ulb.ac.be}
\homepage{http://www.lutsko.com}

\begin{abstract}
The role of metastable liquid phases in vapor-crystal nucleation is studied
using Density Functional Theory(DFT). The bulk free energy functional is
modeled using a hybrid model consisting of a sum of a hard-sphere functional
(Fundamental Measure Theory) and a liquid-like contribution to account for
the attractive part of the interactions. The model gives a
semi-quantitatively accurate description of both the vapor-liquid-solid
phase diagram for both simple fluids (Lennard-Jones interactions) and of the low-density/high-density/crystal phase diagram for model
globular proteins (ten Wolde-Frenkel interaction). The density profile is
characterized by two local order parameters, the average density and the
crystallinity. The bulk free energy model is supplemented by
squared-derivative terms in these order parameters to account for
inhomogeneities thus producing a model similar in spirit to phase-field
theory. It is shown that for both interaction models, the vapor-crystal part
of the phase-diagram can be separated into regions for which metastable
liquid phases are more or less stable than the vapor, but always less stable
than the solid. The former case allows for the possibility of \emph{double
nucleation} whereby liquid droplets nucleate from the vapor followed by a
separate nucleation of the solid phase within the liquid droplets. Whether
or not this actually occurs depends on the relative free energy barriers for
vapor-solid and vapor-liquid nucleation and it is shown that for simple
fluids, double nucleation is indeed favored at sufficiently large
supersaturation. Finally, by studying the minimum free energy path from the
vapor to the solid, the separate possibility of \emph{transient nucleation},
where the vapor-solid transition involves a single nucleation event but the
sub-critical clusters tend to be liquid-like while small, is shown to also
be possible when the metastable liquid exists, but the supersaturation is
too small for double nucleation.
\end{abstract}

\date{\today }
\maketitle
\tableofcontents
\section{Introduction}

The goal of this work is to describe, at a semi-microscopic level, the
process of homogeneous nucleation of solids from solution. This is studied in the approximation that the effects of the solvent can be accounted for by an effective interaction between the solvate molecules so that at low concentrations, the molecules in solution are treated as a single-component gas with the crystalline phase  being the corresponding solid phase in the effective single component system. In this picture, there is also the possibility of a dense liquid-like phase which simply corresponds to the liquid phase of the effective single-component system. Thus, the following work is intended to be applicable both to nucleation in single-component systems, such as simple fluids, and to nucleation of a solid phase from macromolecules in solution. For the sake of consistency, most of the discussion will be phrased in the language of nulceation of solids in a single component system. So, in the following, all statements concerning the ``vapor phase'' should be understood to be equally applicable to the ``low-concentration'' or ``low density'' phase of molecules, such as proteins, in solution and statements concering the ``liquid phase'' are equally applicable to the ``high-concentration'' or ``high-density'' phase of a solution.  

In general, homogeneous
nucleation involves at least two phases: the stable phase, S, and the
metastable phase, M. Initially, the system consists of M in its homogeneous,
bulk state. Fluctuations give rise to small volumes of S which will here
always be taken to be spherical clusters. By definition, the free energy of
bulk S is lower than that of bulk M but finite clusters also involve an
interface which has higher free energy, on average, than either phase.
Hence, sufficiently small clusters are typically unstable, having higher
free energy even than the equivalent amount of M. Furthermore, for small
clusters, there is no reason to believe that the material inside the cluster
is in the bulk S state since the interface can have a volume comparable to
that of the bulk region. Thus, small clusters will be unstable with respect
to M: indeed, will be unstable with respect to smaller clusters thus leading
to cluster dissolution. For sufficiently large clusters, the interior will
consist mostly of material near the bulk S state, and the volume of the
interface will be small compared to the volume of the bulk so that the
cluster is stable with respect to M but unstable with respect to cluster
growth thus driving the transition of the whole system from M to S (see,
e.g. \cite{Kashchiev}).

In recent years, it has become apparent, first for proteins\cite{tWF,
VekilovCGDReview2004, LN, Gunton, GuntonBook} but then even for simple fluids%
\cite{LN, FrenkelLJ, Chen}, that homogeneous nucleation of solids from vapor
can be more complicated than this simple picture. This is because in many
cases, the conditions for solid nucleation also allow for a second
metastable phase, namely that of the bulk liquid. In cases where the bulk
liquid is more stable than the bulk vapor (i.e. has lower free energy) but
less stable than the solid, it can be that, instead of directly nucleating a
crystalline cluster from the vapor, it is energetically favorable to first
nucleate a liquid cluster which then grows and, subsequently, to complete
the transition to the crystalline state via a second homogeneous
(liquid-crystal) nucleation event within the liquid cluster (thus creating a
growing solid cluster within the growing liquid cluster). This process will
be referred to as "double nucleation". Whether or not this actually occurs
will depend on the free energy barrier encountered in the vapor-crystal and
vapor-liquid transition. Alternatively, even if direct vapor-crystal
nucleation is favored, it is still possible that the growing cluster
nevertheless passes through a liquid-like stage which, however, is always
sub-critical, as a transient on the way to creating a solid-like critical
cluster. In this case, small clusters would be liquid-like in structure,
becoming more crystalline as they grow larger. This scenario, involving only
a single nucleation event, will be referred to as "transient two-step
nucleation". The simulations of ten Wolde and Frenkel\cite{tWF} are clear
illustrations of transient two-step nucleation: they exhibit a nucleation
pathway that involves liquid-like small clusters followed by solidification
as the cluster approaches the critical size. Vekilov discusses both
scenarios and suggests that even when double nucleation is possible and
energetically favored, it may be suppressed by kinetics\cite%
{VekilovCGDReview2004}. Kaschiev and Vekilov have analyzed the effect of
double nucleation on observed nucleation rates\cite{KashchievVekilov}. van
Meel et al report simulation results showing double nucleation for a
Lennard-Jones fluid\cite{FrenkelLJ}.

The key to understanding these processes is the construction of models for
the free energy of inhomogeneous - multiphase- systems. Indeed, Classical
nucleation theory (CNT), from which we take our lead, \ is fundamentally
based on a description of the free energy as a function of the size of the
cluster. In CNT\ (which should be thought of in terms of the simpler
vapor-liquid transition) the width of the cluster interface is taken to be
zero and the interior of the cluster is assumed to be in the bulk state, S.
The only variable is then the size of the cluster and it is assumed that
homogeneous nucleation consists of the growth of a cluster from size zero
until it is arbitrarily large\cite{LutskoEPL,Lutsko_JCP_2008_2}. This, in
other words, is understood to be the "nucleation pathway". This idea can be
developed more rigorously to include finite width of the interface and
non-bulk properties of the interior in which case the nucleation pathway
involves a description of the variation of all these properties - size,
interfacial width and interior density - simultaneously\cite{Lutsko_InPrep}.

The process of solid nucleation is more complex, as the phenomenology above
would imply. First, the vapor-solid transition involves at least two order
parameters: the density and the \textit{crystallinity}\cite%
{OLW1,TalOxtoby,Gunton,LN}. Since the typical solid density is close to that
of the metastable liquid (if it exists), the double nucleation scenario
involves a separation of these order parameters: the vapor-liquid transition
involves two different densities but occurs at zero crystallinity while the
liquid-solid transition occurs at (nearly constant)\ density and involves a
change from zero to finite crystallinity. In the transient two-step
scenario, a small (unstable) cluster begins to form at zero crystallinity
but at some point, as the cluster becomes larger, the crystallinity
increases so that the critical cluster has both finite density difference
compared to the vapor and finite crystallinity. Both of these can be viewed
as "two-step" scenarios compared to the possibility that, from the
beginning, the crystallinity and density change at the same time. Clearly,
description of these processes must involve a model for the free energy of
inhomogeneous systems which is sufficiently detailed so as to describe
liquid and solid clusters embedded in a vapor bulk. In order to capture the
transient scenario, this must be supplemented by some means of determining
the nucleation pathway.

Here, the basic tool for calculating the free energies will be classical,
finite-temperature Density Functional Theory (DFT). If DFT were sufficiently
well-developed and technically simple, nothing more would need to be said
about this part of the study: unfortunately, neither of these conditions
holds for the solid phase. (In contrast, direct calculations for
liquid-vapor systems are possible\cite{Lutsko_JCP_2008,Lutsko_JCP_2008_2}.
DFT is only sufficiently well understood so as to give an \textit{a priori}
description of bulk fluid, bulk solid and inhomogeneous fluid-solid systems
for hard-sphere interactions. For any more realistic potential -
particularly those with attractive interactions - some sort of modelling is
necessary. Typically, this will involve the introduction of an effective
hard-sphere diameter and the representation of the free energy as the sum of
the hard-sphere functional and a second term accounting for the attractive
part of the interactions. Futherrmore, even for the hard-sphere functional,
the calculation of the free energy for an inhomogeneous system (i.e. of a
solid cluster embedded in a fluid background) is very complicated and
computationally expensive\cite{Song_FMT}. Fortunately, an easier alternative
is available. It has been shown that the exact free energy for a solid can
be expanded in terms of gradients of the order parameters\cite%
{Lowen1,Lowen2,LutskoGL} thus providing a connection between DFT and the
older gradient theories of Cahn et al\cite%
{CahnHilliard,CahnHilliardNucleation}, as well as the phase-field theories
commonly used to study solid-solid transitions\cite{phase_field}. This is
the justification for using a gradient model in the present work. A
significant advantage is that the free energy functional is then only needed
for the case of homogeneous (i.e. bulk) systems, thus placing less stress on
the accuracry of the DFT model. On the other hand, expressions for the
coefficients of the gradient terms must be found. In principle, the exact
results express these in terms of derivatives of the bulk free energy but in
practice, they are hard to calculate except for the case of a fluid. In this
work, a semi-empirical procedure will be used to fix these terms. The main
difference between the present work and that of Gunton et al\cite{Gunton},
which is similar in spirit, is that the latter made use of toy free energy
models whereas here realistic models are used which give semi-quantitatively
accurate bulk phase diagrams as well as of the liquid-vapor phase transition%
\cite{Lutsko_JCP_2008}.

Given these approximations, the free energy for any configuration of order
parameters can be calculated. The practical exploration of the models will
make use of energy-surface techniques commonly applied to the study of
chemical reaction pathways and structural transitions\cite{Wales}. A
recently developed method - involving the approximation of the
spatially-varying order parameters as piecewise-continuous functions, will
be used to determine the critical clusters - i.e. saddle points of the free
energy - for homogeneous nucleation. This will already give enough
information to identify if and when double nucleation is possible. The
nucleation pathway will be identified, as is commonly done for chemical
reaction pathways, using steepest descent paths. These identify the most
likely path for the transition given the free energy surface and are a
natural generalization of the CNT pathway. They do not take into account
kinetic factors, such as rates of mass transport, that could play a
significant role particularly for small molecules.

Section II will describe the technical details of the free energy models
used. The bulk thermodynamics is used in Section III to limit the regions of
the phase diagram in which double nucleation is possible. A simple model for
double nucleation is also used to illustrate the role of bulk free energy
differences and of surface tension. The fourth Section describes the
application of the model to planar interfaces and illustrates the role of
wetting. Detailed calculations of the energy barriers for direct nucleation
of the crystal and of double nucleation are presented in Section V. The
possibility of transient double nucleation is also described. The paper ends
with a summary and with a discussion of future directions.

\section{The free energy model}

\subsection{Density}

The central quantity defining the system state in DFT is the local density, $%
\rho(\mathbf{r})$. It is possible to formulate the theory with no a priori
restrictions placed on the density, but this is computationally expensive. A
commonly used alternative is to represent the density in terms of a set of
basis functions. For bulk crystalline systems, this usually means a sum of
Gaussians localized at the lattice sites, 
\begin{equation}
\rho\left( \mathbf{r};x,\rho_{latt},\alpha\right) =x\sum_{i}\left( \frac{%
\alpha}{\pi}\right) ^{3/2}\exp\left( -\alpha\left( \mathbf{r}-\mathbf{R}%
_{i}\left( \rho_{latt}\right) \right) ^{2}\right)
\end{equation}
where $\mathbf{R}_i$ is a lattice vector, $\alpha$ controls the width, $x$
is the occupancy and $\rho_{latt}$ is the lattice density. The average
density is 
\begin{equation}
\overline{\rho}=\frac{1}{V}\int_{V}\rho\left( \mathbf{r};x,\rho_{latt},%
\alpha\right) =x\rho_{latt}
\end{equation}
The density can also be written in Fourier representation as%
\begin{equation}
\rho\left( \mathbf{r};x,\rho_{latt},\alpha\right)
=x\rho_{latt}\sum_{i}\exp\left( i\mathbf{r}\cdot\mathbf{K}_{i}\left(
\rho_{latt}\right) \right) \exp\left( -\mathbf{K}_{i}^{2}\left(
\rho_{latt}\right) /4\alpha\right)
\end{equation}
where $\mathbf{K}_i$ is a reciprocal lattice vector. Notice that in this
representation, it is clear that the limit $\alpha \rightarrow 0$ gives a
uniform density $\rho(\mathbf{r}) = \overline{\rho}$ which describes a
fluid. It is therefore natural to characterize the crystallinity by the size
of the amplitude of a typical nonzero wavevector term such as 
\begin{equation*}
\chi=\exp\left( -\mathbf{K}_{1}^{2}\left( \rho_{latt}\right) /4\alpha \right)
\end{equation*}
so that the density can also be written as%
\begin{equation}
\rho\left( \mathbf{r};x,\rho_{latt},\alpha\right) =x\rho_{latt}\sum
_{i=0}\exp\left( i\mathbf{r}\cdot\mathbf{K}_{i}\left( \rho_{latt}\right)
\right) \chi^{\left( K_{i}\left( \rho_{latt}\right) /K_{1}\left(
\rho_{latt}\right) \right) ^{2}}
\end{equation}
thus showing that the density is parameterized entirely by $x$, $\rho_{latt}$
and $\chi$. In the following, we neglect the variation of the lattice
density and generalize to inhomogeneous systems by allowing the occupancy
and the crystallinity to depend on position. It is also convenient then to
replace the occupancy by the average density so that we have%
\begin{equation}
\rho\left( \mathbf{r}\right) =\overline{\rho}\left( \mathbf{r}\right)
\sum_{i=0}\exp\left( i\mathbf{r}\cdot\mathbf{K}_{i}\left( \rho
_{latt}\right) \right) \left( \chi\left( \mathbf{r}\right) \right) ^{\left(
K_{i}\left( \rho_{latt}\right) /K_{1}\left( \rho_{latt}\right) \right) ^{2}}
\end{equation}
The order parameters are then the local average density, $\overline{\rho}%
\left( \mathbf{r}\right) $, and the crystallinity, $\chi\left( \mathbf{r}%
\right) $. It will sometimes be more convenient to replace the latter by the
amplitude of the smallest non-zero wavevector, $\rho_{1}(\mathbf{r}) = 
\overline{\rho}\left( \mathbf{r}\right)\chi\left( \mathbf{r}\right)$.

\subsection{Gradient expansion}

In order to determine the density, a model for the (grand canonical) free
energy functional, $\Omega \lbrack \rho ]$ is necessary. Good models exist
for liquids and can be used to study, e.g., the liquid-vapor transition in
great detail\cite{Lutsko_JCP_2008}. However, the theory is less well
developed for the solid phase and in any case calculations for inhomogeneous
solids are very expensive. The present work therefore makes use of a
gradient expansion of the free energy which focusses attention on the order
parameters and only requires information about homogeneous solids\cite%
{Lowen1,Lowen2,LutskoGL}. The grand-canonical free energy is written as 
\begin{equation}
\Omega \lbrack \rho ]=F[\rho ]-\mu \int \rho (\mathbf{r})d\mathbf{r}
\end{equation}%
where $\mu $ is the chemical potential. In general, if the density can be
written in terms of $n$ order parameters, $\Gamma =\left\{ \Gamma
_{1},...,\Gamma _{n}\right\} ,$ as 
\begin{equation}
\rho \left( \mathbf{r}\right) =f\left( \mathbf{r};\Gamma \left( \mathbf{r}%
\right) \right)
\end{equation}%
so that the density of the uniform, bulk system is 
\begin{equation}
\rho _{\Gamma }\left( \mathbf{r}\right) =f\left( \mathbf{r};\Gamma \right)
\end{equation}%
and if $\Gamma \left( \mathbf{r}\right) $ in some sense "slowly varying",
then the squared-gradient approximation (SGA) for the functional $F[\rho ]$
is 
\begin{equation}
F\left[ \rho \right] \simeq \int \left\{ f\left( \Gamma \left( \mathbf{r}%
\right) \right) +\frac{1}{2}K_{ab}^{ij}\left( \Gamma \left( \mathbf{r}%
\right) \right) \frac{\partial \Gamma _{i}\left( \mathbf{r}\right) }{%
\partial r_{a}}\frac{\partial \Gamma _{j}\left( \mathbf{r}\right) }{\partial
r_{b}}\right\}
\end{equation}%
where the summation convention is used. The first term of the free energy
involves the bulk free energy density defined as 
\begin{equation}
f\left( \Gamma \right) =\frac{1}{V}F\left[ \rho _{\Gamma }\right]
\end{equation}%
For the order parameters used here, the free energy is explicitly 
\begin{equation}
F\left[ \rho \right] \simeq \int \left\{ 
\begin{array}{c}
f\left( \overline{\rho }\left( \mathbf{r}\right) ,\chi \left( \mathbf{r}%
\right) \right) +\frac{1}{2}K_{ab}^{\rho \rho }\left( \overline{\rho }\left( 
\mathbf{r}\right) ,\chi \left( \mathbf{r}\right) \right) \frac{\partial 
\overline{\rho }\left( \mathbf{r}\right) }{\partial r_{a}}\frac{\partial 
\overline{\rho }\left( \mathbf{r}\right) }{\partial r_{b}} \\ 
+K_{ab}^{\rho \chi }\left( \overline{\rho }\left( \mathbf{r}\right) ,\chi
\left( \mathbf{r}\right) \right) \frac{\partial \overline{\rho }\left( 
\mathbf{r}\right) }{\partial r_{a}}\frac{\partial \chi \left( \mathbf{r}%
\right) }{\partial r_{b}}+\frac{1}{2}K_{ab}^{\chi \chi }\left( \overline{%
\rho }\left( \mathbf{r}\right) ,\chi \left( \mathbf{r}\right) \right) \frac{%
\partial \chi \left( \mathbf{r}\right) }{\partial r_{a}}\frac{\partial \chi
\left( \mathbf{r}\right) }{\partial r_{b}}%
\end{array}%
\right\}
\end{equation}

\subsection{Bulk Free Energy}

The bulk free energy model used here is based on the idea of separating the
free energy into a hard-sphere contribution, for which the DFT is well
developed, and a second contribution that accounts for the long-ranged
attractive interactions. A particularly simple model is based on the
observation that the local structure of an FCC solid and a simple fluid are
quite similar so that, as a simplest approximation, one can imagine that the
correction to the hard-sphere model is independent of the local structure%
\cite{Curtin, LN} giving 
\begin{equation}
\frac{1}{V}F\left[ \rho_{\Gamma}\right] =\frac{1}{V}F_{HS}\left[
\rho_{\Gamma}d\left( \Gamma\right) ^{3}\right] +f_{tail}\left(
\rho_{\Gamma},d\left( \Gamma\right) \right)
\end{equation}
where $d\left( \Gamma\right) =d\left( \overline{\rho}\right) $ is the
effective hard-sphere diameter and 
\begin{equation}
f_{tail}\left( \rho_{\Gamma},d\left( \Gamma\right) \right) =\frac{1}{V}F%
\left[ \overline{\rho}\right] -\frac{1}{V}F_{HS}\left[ \overline{\rho }d ^{3}%
\right]
\end{equation}
is the contribution of the attractive part of the interaction to the liquid
free energy. The effective hard-sphere diameter is determine using the WCA
expression as modified by Ree et al\cite{Ree1,Ree2}.

\subsection{Bulk phase diagram}

In DFT, one is always working in the grand-canonical ensemble so the
external parameters are the temperature, the chemical potential, $\mu$ and
the applied external field, $\phi\left( \mathbf{r}\right) $. The latter
includes any confining walls:\ if the walls are hard, then the volume is a
fixed parameter (as will always be assumed here). The appropriate free
energy is the grand potential,%
\begin{equation*}
\Omega\left[ \rho\right] =F\left[ \rho\right] -\mu\int\rho\left( \mathbf{r}%
\right) d\mathbf{r+}\int\phi\left( \mathbf{r}\right) \rho\left( \mathbf{r}%
\right) d\mathbf{r}
\end{equation*}
and it should be noted that all information about the state is encoded in
the local density function, $\rho\left( \mathbf{r}\right) $. The equilibrium
states (i.e. density distributions) are determined by minimization,%
\begin{equation}
0=\frac{\delta\Omega\left[ \rho\right] }{\delta\rho\left( \mathbf{r}\right) }
\end{equation}
which is to say%
\begin{equation}
\frac{\delta F\left[ \rho\right] }{\delta\rho\left( \mathbf{r}\right) }%
=\mu-\phi\left( \mathbf{r}\right)
\end{equation}
For parameterized profiles, the requirement that the free energy be a
minimum gives%
\begin{equation}
0=\frac{\delta\Omega\left[ \rho_{\Gamma}\right] }{\delta\Gamma\left( \mathbf{%
r}\right) }
\end{equation}
and if the parameters are constants, then%
\begin{equation}
0=\frac{\partial\Omega\left[ \rho_{\Gamma}\right] }{\partial\Gamma}
\end{equation}

For a given value of chemical potential, there may be multiple solutions for
the density in which case the equilibrium state is the one corresponding to
the absolute minimum of the grand potential. Two phase coexistence therefore
requires that%
\begin{align}
\left. \frac{\delta F\left[ \rho\right] }{\delta\rho\left( \mathbf{r}\right) 
}\right\vert _{\rho_{1}} & =\mu-\phi\left( \mathbf{r}\right) =\left. \frac{%
\delta F\left[ \rho\right] }{\delta\rho\left( \mathbf{r}\right) }\right\vert
_{\rho_{2}} \\
\Omega\left[ \rho_{1}\right] & =\Omega\left[ \rho_{2}\right]  \notag
\end{align}
In particular, using the chain rule for functional differentiation, 
\begin{align}
\frac{\partial F\left[ \rho\right] }{\partial\overline{\rho}} & =\int\frac{%
\delta F\left[ \rho\right] }{\delta\rho\left( \mathbf{r}\right) }\frac{%
\partial\rho\left( \mathbf{r}\right) }{\partial\overline{\rho}}d\mathbf{r} \\
& =\int\frac{\delta F\left[ \rho\right] }{\delta\rho\left( \mathbf{r}\right) 
}d\mathbf{r}  \notag
\end{align}
gives the usual thermodynamic relation%
\begin{equation}
\frac{\partial\frac{1}{V}F\left[ \rho\right] }{\partial\overline{\rho}}=\mu-%
\frac{1}{V}\int\phi\left( \mathbf{r}\right) d\mathbf{r.}
\end{equation}
Thus, for uniform densities, $\rho_{i}\left( \mathbf{r}\right) =\overline{%
\rho}_{i}$ and e.g. $F\left[ \rho_{i}\right] \rightarrow F\left( \overline{%
\rho}_{i}\right) $, the conditions for coexistence are 
\begin{align}
\frac{\partial\frac{1}{V}F\left( \overline{\rho}_{1}\right) }{\partial 
\overline{\rho}_{1}} & =\mu-\frac{1}{V}\int\phi\left( \mathbf{r}\right) d%
\mathbf{r}=\frac{\partial\frac{1}{V}F\left( \overline{\rho}_{2}\right) }{%
\partial\overline{\rho}_{2}}  \label{cx} \\
\Omega\left( \rho_{1}\right) & =\Omega\left( \rho_{2}\right)  \notag
\end{align}
which is to say equality of chemical potentials and of pressures since in
the bulk, $\Omega=-PV$.

For the crystalline system, the parameterization used here involves not just
the average density but also the crystallinity and the lattice parameter.
The conditions for an extremum of the free energy are then 
\begin{align}
\frac{\partial \frac{1}{V}F\left( \overline{\rho },\chi ,\rho _{latt}\right) 
}{\partial \overline{\rho }}& =\mu -\frac{1}{V}\frac{\partial }{\partial 
\overline{\rho }}\int \rho \left( \mathbf{r};\overline{\rho },\chi ,\rho
_{latt}\right) \phi \left( \mathbf{r}\right) d\mathbf{r} \\
\frac{\partial \frac{1}{V}F\left( \overline{\rho },\chi ,\rho _{latt}\right) 
}{\partial \chi }& =-\frac{1}{V}\frac{\partial }{\partial \chi }\int \rho
\left( \mathbf{r};\overline{\rho },\chi ,\rho _{latt}\right) \phi \left( 
\mathbf{r}\right) d\mathbf{r}  \notag \\
\frac{\partial \frac{1}{V}F\left( \overline{\rho },\chi ,\rho _{latt}\right) 
}{\partial \rho _{latt}}& =-\frac{1}{V}\frac{\partial }{\partial \rho _{latt}%
}\int \rho \left( \mathbf{r};\overline{\rho },\chi ,\rho _{latt}\right) \phi
\left( \mathbf{r}\right) d\mathbf{r}  \notag
\end{align}%
In principle, the chemical potential (an external field) are specified and
these equations solved for $\overline{\rho },\chi $ and $\rho _{latt}$:\ in
practice, it is simpler to choose a value of the lattice density and to use
these conditions to determine $\overline{\rho },\chi $ and $\mu $. Further
technical details are discussed in Appendix \ref{Bulk}.

\section{Thermodynamics of two step nucleation}

\subsection{Independent variables and Ensembles}

The main calculational tool used here, DFT, is formulated in the
grand-canonical ensemble in which the independent variables are the
temperature, chemical potential and volume (or, more generally, applied
field) and the free energy of interest is the grand potential. Experiments
are typically performed at constant pressure, temperature and volume for
which the Gibbs free energy is relevant. Fortunately, in discussing
nucleation, one is always interested in free energy differences and these
are independent of the ensemble\cite{Oxtoby_Evans}. In the following, since
it is based on DFT, the grand-canonical ensemble is always used, however in
many cases the difference between ensembles can be suppressed by focussing
on physical quantities. Thus, rather than specify the chemical potential
(for the grand ensemble) or the pressure (for the PVT ensemble) one can
specify the density of the final phase which is a meaningful variable in
both formulations and in either case is uniquely related to the state
variable.

\subsection{Interaction potentials and the fluid phase}

In this study, the nucleation properties of two different systems will be
considered:\ simple fluids as described by the Lennard-Jones potential,%
\begin{equation}
v_{LJ}\left( r\right) =4\epsilon\left( \left( \frac{\sigma}{r}\right)
^{12}-\left( \frac{\sigma}{r}\right) ^{6}\right)
\end{equation}
and globular proteins as described by the ten Wolde-Frenkel model
interaction,%
\begin{equation}
v_{tWF}(r)\,=\left\{ 
\begin{array}{c}
\infty,\;r>\sigma \\ 
\,\frac{4\,\epsilon}{\alpha^{2}}\left( \,\left( \frac{1}{(\frac{r}{\sigma }%
)^{2}-1}\right) ^{6}-\,\alpha\,\left( \frac{1}{(\frac{r}{\sigma})^{2}-1}%
\right) ^{3}\right) ,\;r\geq\sigma%
\end{array}
\right. ,
\end{equation}
which will be studied here for the value $\alpha=50$ as is appropriate to
model the phase behavior of globular proteins.

Both of these potentials are long-ranged in the sense of decaying as
power-laws. In simulation, infinite-ranged potentials are difficult to deal
with, so any long-ranged potential $v(r)$ is typically cutoff at some
distance, $r_{c}$. For Monte-Carlo, a simple shift of the potential to make
the energy continuous at the cutoff is typically used so that the so-called
truncated and shifted potential is%
\begin{equation}
v_{mc}\left( r\right) =\left\{ 
\begin{array}{c}
v(r)-v\left( r_{c}\right) ,\;r\leq r_{c} \\ 
0,\;r>r_{c}%
\end{array}%
\right.
\end{equation}%
In molecular dynamics simulations, the force is usually required to be
continuous so that the force-shifted potential is commonly used,%
\begin{equation}
v_{md}\left( r\right) =\left\{ 
\begin{array}{c}
v(r)-v\left( r_{c}\right) -v^{\prime }\left( r_{c}\right) \left(
r-r_{c}\right) ,\;r\leq r_{c} \\ 
0,\;r>r_{c}%
\end{array}%
\right.
\end{equation}%
where $v^{\prime }(r)=dv(r)/dr$. Other forms are also important,
particularly the Broughton-Gilmer modification of the LJ potential:%
\begin{equation}
v_{BG}\left( r\right) =\left\{ 
\begin{array}{c}
v_{LJ}(r)+C_{1},\;r\leq 2.3\sigma \\ 
C_{2}\left( \frac{\sigma }{r}\right) ^{12}+C_{3}\left( \frac{\sigma }{r}%
\right) ^{6}+C_{4}\left( \frac{\sigma }{r}\right) ^{2}+C_{5},\;2.3\sigma
<r\leq 2.5\sigma \\ 
0,\;r>2.5\sigma%
\end{array}%
\right.
\end{equation}%
where $C_{1}=0.016132\epsilon $, $C_{2}=313.66\epsilon $, $%
C_{3}=-68.069\epsilon $, $C_{4}=0.083312\epsilon $ and $C_{5}=0.74689%
\epsilon $\cite{BG1}.

All of these details significantly affect the liquid-vapor phase diagram.
The DFT model described above requires as input both the interaction
potential and the equation of state for the fluid phase. For the LJ
potential with \emph{no cutoff}, essentially exact empirical equations of
state are available for temperatures above the triple point\cite{JZG,Mecke, Mecke1}.
For finite cutoffs, these must be modified with inexact mean-field
corrections. Although useful for studying liquid-vapor coexistence\cite%
{Lutsko_JCP_2008,Lutsko_JCP_2008_2}, they are of limited utility for
vapor-solid nucleation since the interesting affects occur below the triple
point. Less accurate, but more broadly applicable, are mean-field equations
of state based on thermodynamic perturbation theory such as the
Weeks-Chandler-Andersen (WCA) theory\cite{WCA1,WCA2,WCA3,Ree1,Ree2} which
will be used here.

\subsection{Solid phase diagrams and intermediate phases}

\begin{figure*}[tbp]
\includegraphics[angle=-0,scale=0.4]{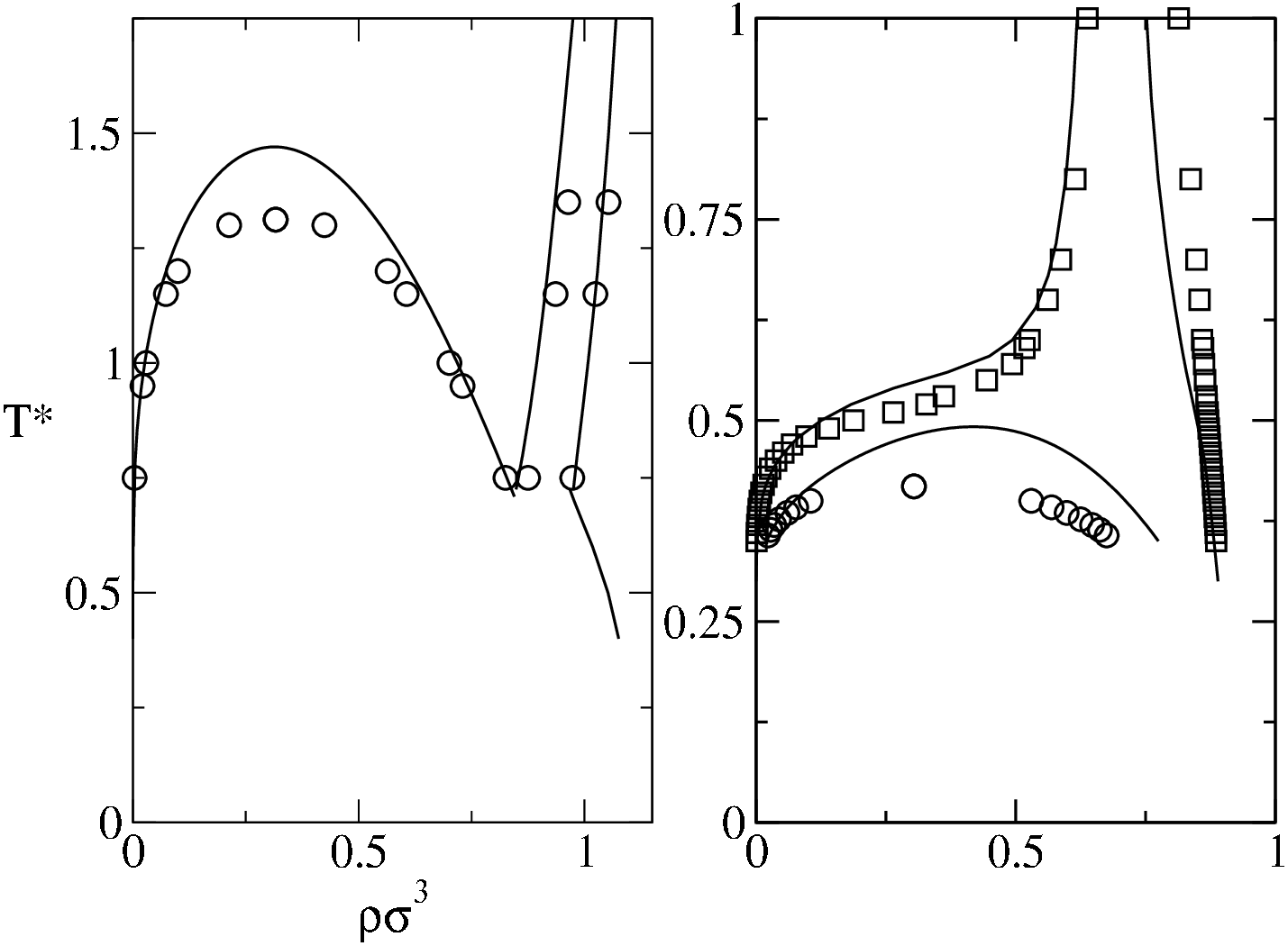} %%{phase_diagrams}
\caption{Phase diagrams for the Lennard-Jones potential, left panel, and the
tWF potential, right panel, together with simulation data. The tWF potential
was cutoff and shifted at $r=2.8\protect\sigma$.}
\label{figPhase1}
\end{figure*}
Figure \ref{figPhase1} shows the calculated vapor-liquid-FCC solid phase
diagram for the infinite-ranged LJ potential and for the truncated and
shifted tWF potential compared to simulation in terms of the reduced
temperature $T^{*}=k_BT/\epsilon$. The LJ phase diagram possesses both a
liquid-vapor critical point and a triple point whereas for the protein
model, the liquid-vapor transition is metastable. Since the fluid equation
of state is being calculated from thermodynamic perturbation theory, which
is a mean field theory, the critical point is, as expected, poorly described
for both potentials. Apart from this expected inaccuracy, the model is in
semi-quantitative agreement with simulation for all three phases.

\subsubsection{Nucleation scenarios for globular proteins}

\begin{figure*}[b]
\includegraphics[angle=-0,scale=0.4]{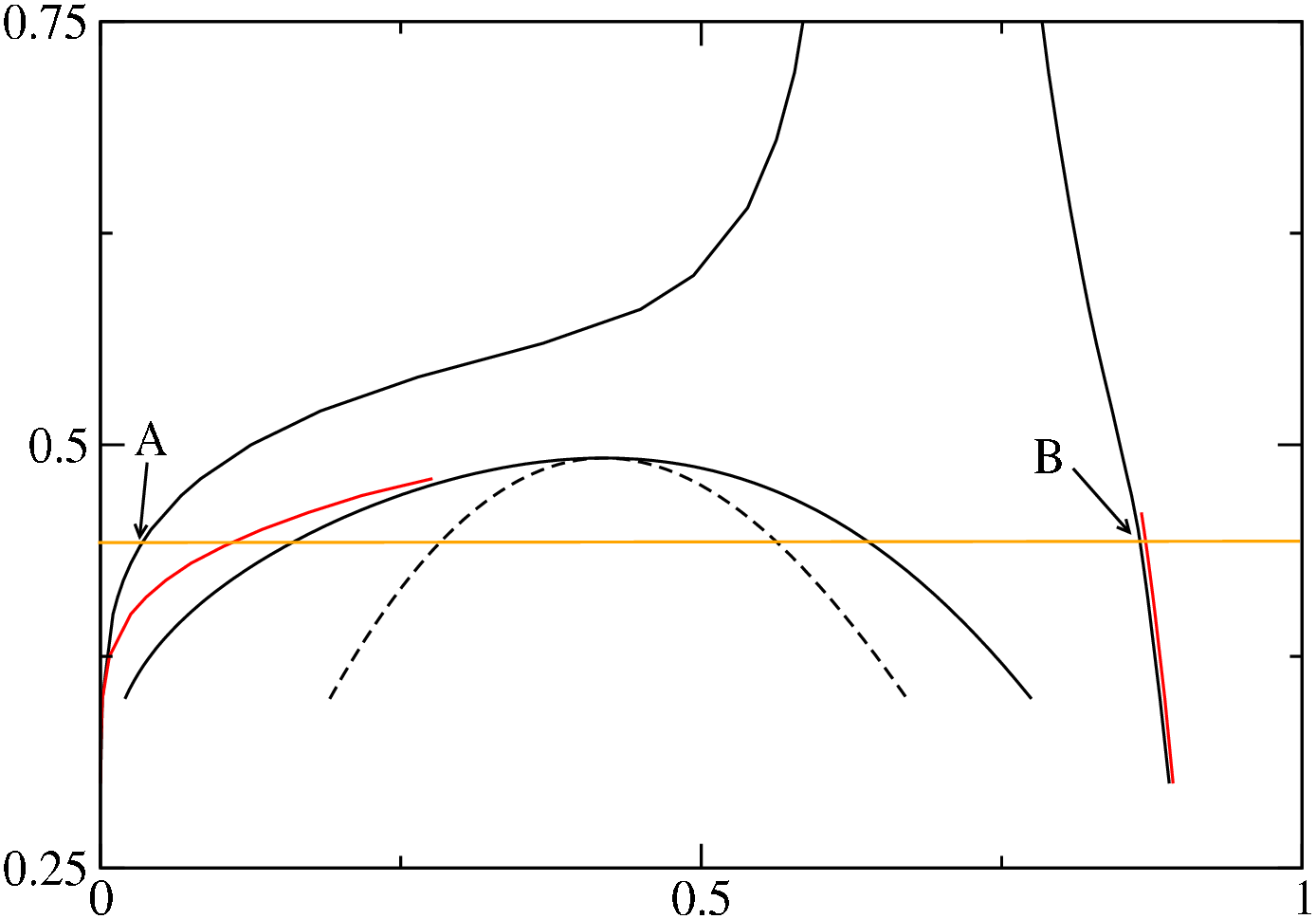} %%{phase_diagrams_no_data_twf}
\caption{The phase diagram for the tWF potential as in Fig. \protect\ref{figPhase1} but without the simulation
data and showing the spinodal for the vapor-liquid transition (broken line). The figure also shows, in red, the boundaries for the existence of the
intermediate liquid phase: there is no liquid phase for chemical potentials corresponding to vapor densities to the left of the red line. A similar line for the solid phase is nearly indistinguishable from the solid binodal. Similar boundaries also exist on the Lennard-Jones phase diagram but are so close to the binodals as to be almost indistinguishable. The horizontal line, an isotherm, picks out coexisting states with the coexisting vapor and solid states being labeled A and B, respectively.}
\label{figPhase2}
\end{figure*}
In order to clarify the thermodynamics of metastable states, we consider in
more detail the phase diagram derived from the tWF potential. Figure \ref%
{figPhase2} shows a line crossing the phase diagram at constant temperature.
The points A and B identify coexisting vapor and solid phases which, by
definition, have the same free energy and the same control parameter
(chemical potential or pressure). Starting at the vapor point A and moving along the isotherm in the direction of increasing density, i.e.
to the right, corresponds to increasing the chemical potential and to each value of the density
 there will be a point on the isotherm to the right of the solid-branch of the binodal, i.e. to the right of point B,  having the same chemical potential; however, under these conditions the solid phase will have lower free
energy than the vapor phase so that the vapor points to the right of A are 
\emph{metastable} with respect to the solid phase. Eventually, as one moves along the isotherm, the point in the vapor region reaches the vapor-liquid coexistence
curve so that there is a coexisting liquid phase. However, by definition, this particular liquid state has the same free energy as the vapor, one knows that it has higher free energy than the solid and so can only be metastable. The remains true as one moves along the isotherm to densities to the right of the vapor branch of the liquid-vapor binodal.  Eventually, one reaches the vapor-liquid spinodal and above this density, the vapor phase does not exist. The set of liquid points with the same chemical potential as the vapor points on the spinodal will therefore divide region to the right of the liquid binodal into a lower-density region, for which a vapor with the same chemical potential can always be found, and a higher-density region with no corresponding vapor. 

What is the role of the metastable liquid phase in vapor-solid nucleation?\
We can only address this question here with respect to double nucleation.
(Transient metastable states are a property of the nucleation pathway and
its presence or absence cannot be answered solely from a consideration of
the bulk phases.) First, consider again the vapor-liquid coexistence curve.
By a reversal of the reasoning above, vapor states to the left of the vapor
branch correspond to liquid points to the left of the liquid-branch of the
coexistence curve and are more stable than the corresponding liquid state.
Moving to the left away from the vapor branch, the corresponding liquid
point also moves left until it reaches the liquid-vapor spinodal: beyond
this point, there is no corresponding liquid. This therefore defines a line
in the phase diagram with the property that vapor states to the left have no
corresponding liquid states, shown as a red line in Fig. \ref{figPhase2}. In
particular, points on the vapor branch of the vapor-solid coexistence curve
have no corresponding liquid. This line therefore divides the metastable vapor region into two parts: a low density part in which there is no corresponding liquid state (i.e. no liquid with the same chemical potential) that can play a role in nucleation and a higher density region where the corresponding liquid exists as a metastable state. Hence, \emph{for systems prepared with the
vapor density between the vapor-branch of the solid-vapor coexistence curve
and the new demarcation line, double nucleation is not possible as there is
no metastable liquid. } The region within which vapor and liquid states with the same chemical potential can be found is therefore an envelope around the binodal and will be referred to as the $\mu$-nodal curve. 

Starting at the vapor branch of the vapor-solid coexistence curves, and
moving to the right we therefore find that:

\begin{enumerate}
\item From coexistence to the $\mu$-nodal line, there is no liquid phase
and so, double nucleation is not possible. In this region, the solid free
energy is less than that of the vapor.

\item From the $\mu$-nodal line, to the vapor branch of the vapor-liquid
coexistence curves, there is a liquid state but it has higher free energy
than the vapor (which has higher free energy than the solid). Nucleation via
a liquid cluster is possible, with the solid cluster nucleating within the
liquid but the liquid cluster would always be metastable. This is a form of
transient two-step nucleation.

\item From the vapor branch of the liquid-vapor coexistence curve to the
vapor-liquid spinodal, the liquid has lower free energy than the vapor but
higher than the solid. True double nucleation is possible depending on the
barriers for the formation of a critical liquid cluster within the vapor, a
critical solid cluster in the liquid as compared to the barrier for directly
forming a critical solid cluster in the vapor. Even if the latter is
favored, a transient scenario is possible.
\end{enumerate}

\begin{figure*}[tbp]
\includegraphics[angle=-0,scale=0.4]{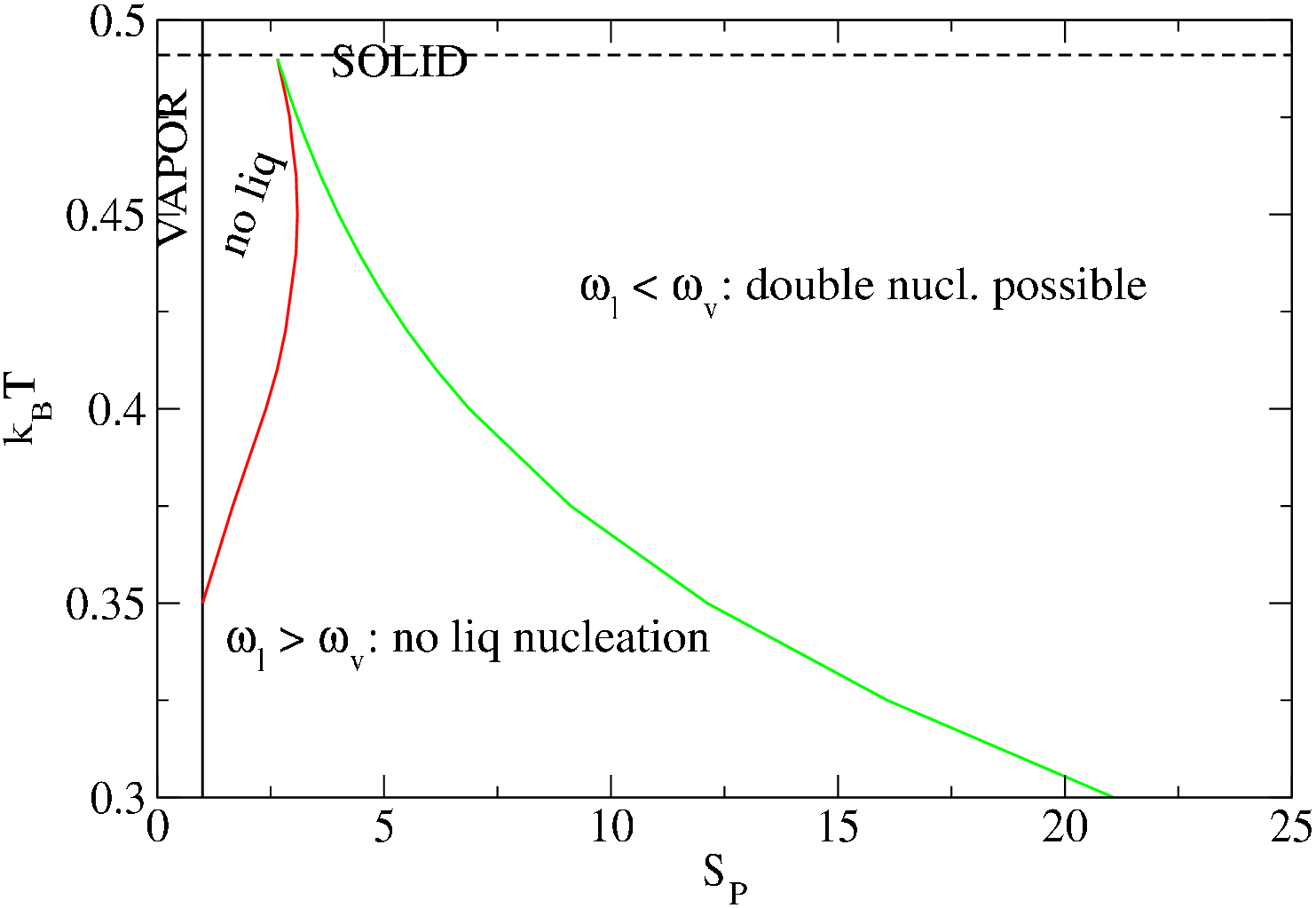} %%{S-kT-tWF}
\caption{Phase diagrams for the tWF potential with (vapor-solid)
supersaturation as the independent variable.}
\label{tWF-S}
\end{figure*}

All of this is shown in a more direct way for the tWF potential in Fig. \ref%
{tWF-S} where supersaturation, $S = P_{v}/P_{coex}$ with $P_{v}$ the vapor
pressure and $P_{coex}$ the vapor-solid coexistence pressure, is used as the
independent variable. By definition, for $S<1$, the vapor is the stable
phase and for $S>1$ the solid is the stable phase. The latter region is
divided into three sections: that for which there is no corresponding liquid
state, that in which there is a liquid but it has higher free energy than
the vapor and that in which there is a liquid state with lower free energy
than the vapor but higher free energy than the solid. Only in the latter
case is liquid nucleation possible so it is only in this region of the phase
diagram that double nucleation can occur. This therefore represents a set of
necessary conditions for double nucleation. Nucleation pathways that do not
involve double nucleation but which pass through liquid-like state, i.e.
transient two-step nucleation, can in principle occur for any value of $S>1$%
. The primary goal of detailed DFT calculations is to determine the
necessary conditions for double nucleation and to assess for what
conditions, if any, transient two-step nucleation occurs.

\subsubsection{Nucleation scenarios for simple fluids}

\begin{figure*}[tbp]
\includegraphics[angle=-0,scale=0.4]{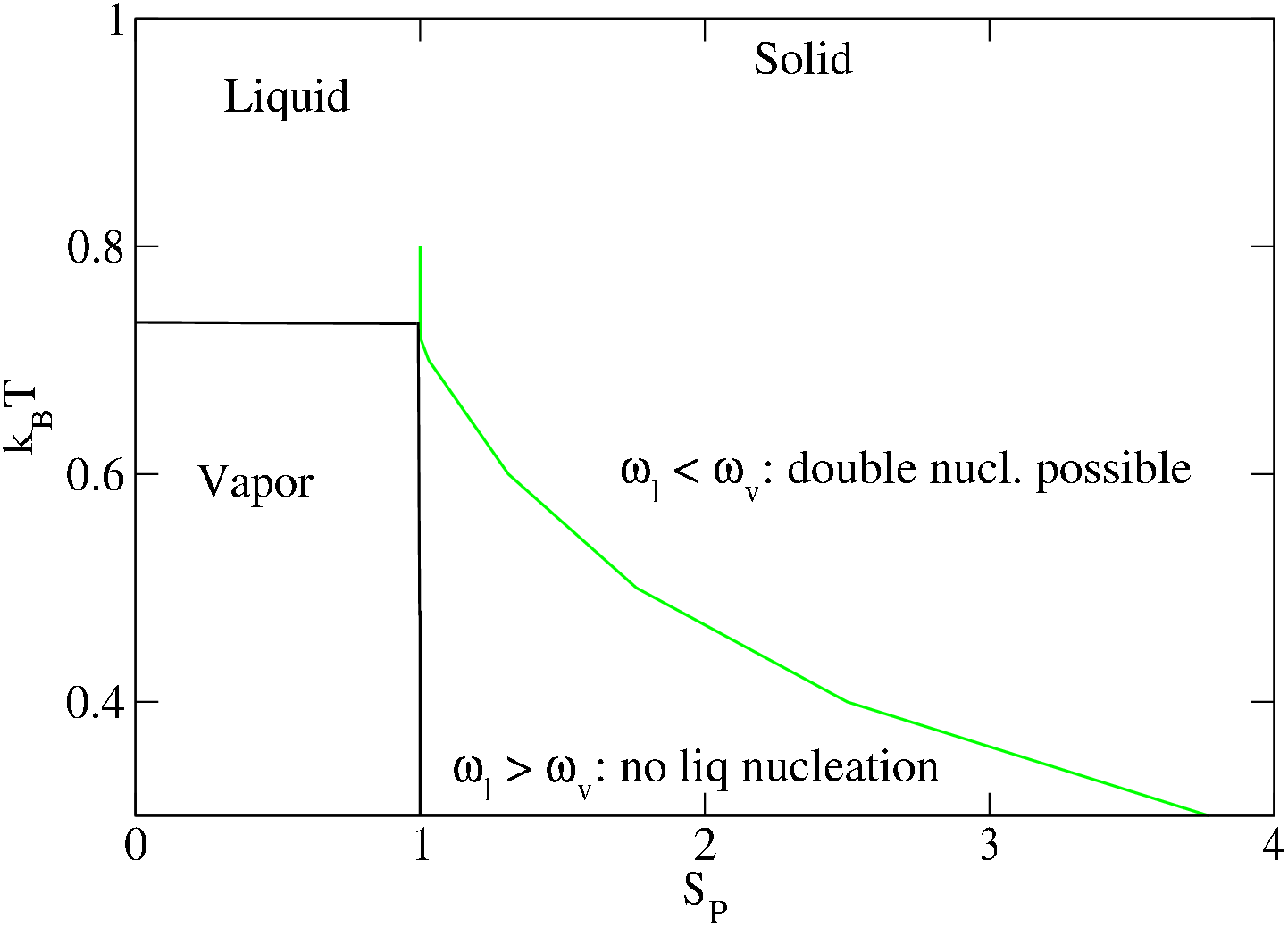} %%{S-kT-LJ}
\caption{Phase diagram for the LJ potential with (vapor-solid)
supersaturation as the independent variable.}
\label{LJ-S}
\end{figure*}

In some ways, the phase diagram for simple liquids is more complex because
the various coexistence curves cross. As shown in Fig.\ref{figPhase2}, the
calculations indicate that there is liquid-vapor coexistence below the
triple point. However, just as in the case of the globular proteins, the
vapor branch of the vapor-liquid coexistence curve is to the right of the
vapor branch of the vapor-solid coexistence curves thus indicating that the
coexisting liquid and vapor have higher free energy than does the solid
phase (at the value of chemical potential or pressure corresponding to
liquid-vapor coexistence). Hence, below the triple point, the liquid phase
is again metastable just as in the case of the globular proteins. Similarly,
below the triple point the liquid branch of the liquid-solid coexistence
curves lies to the left of the liquid branch of the liquid-vapor coexistence
curve: i.e. it is in the metastable (or even unstable) region of the
liquid-vapor phase diagram. So, these liquid states have higher free energy
than the corresponding vapor phase and the coexisting liquid and solid
phases are metastable with respect to the vapor. Taking all of this
together, only the vapor-solid transition is usually drawn below the triple
point, but there is a metastable liquid and a liquid-vapor coexistence curve
in this region. It is this metastable liquid phase which gives rise to the
possibility of two-step nucleation even for simple liquids. However, the
fact that the vapor branches of the vapor-solid coexistence curves and the
vapor-liquid coexistence curves are very close together makes it more
difficult to display the various metastability boundaries. Figure \ref{LJ-S}
shows the phase information in terms of the supersaturation and the
similarities and differences to the model protein behavior are evident. The
main difference is that the intermediate liquid state exists for all values
of the supersaturation leaving only the two regions distinguishing liquids
which are less or more stable than the vapor phase.

\subsection{A simple picture of double nucleation}

In this subsection, the goal is to anticipate the following, more technical,
developments to give an idea of how double-nucleation can be described by
something that looks like an extension of CNT. We begin with the free energy
functional defined above specialized to spherical symmetry, 
\begin{equation}
\Omega \left[ \rho \right] \simeq \int \left\{ \omega \left( \overline{\rho }%
\left( r\right) ,\chi \left( r\right) \right) +\frac{1}{2}K^{\rho \rho
}\left( \frac{\partial \overline{\rho }\left( r\right) }{\partial r}\right)
^{2}+K^{\rho \chi }\frac{\partial \overline{\rho }\left( r\right) }{\partial
r}\frac{\partial \chi \left( r\right) }{\partial r}+\frac{1}{2}K^{\chi \chi
}\left( \frac{\partial \chi \left( r\right) }{\partial r}\right)
^{2}\right\} 
\end{equation}%
where $\omega =f-\mu \rho $ is the grand potential per unit volume and the
coefficients of the gradient terms are taken to be constants. It is assumed
that the temperature and chemical potential are such that there are three
bulk states:\ a solid with order parameters $\left( \overline{\rho }%
_{s},\chi _{s}\right) $, a liquid with order parameters $\left( \overline{%
\rho }_{l},0\right) $ and a vapor with parameters $\left( \overline{\rho }%
_{v},0\right) $.We now introduce a very simple model for the order
parameters for a (spherically symmetric) vapor-solid cluster which
nevertheless captures the important physics of the real system:%
\begin{align}
\overline{\rho }\left( \mathbf{r}\right) & =\left\{ 
\begin{array}{c}
\overline{\rho }_{s},\;r<R \\ 
\overline{\rho }_{s}+\left( \overline{\rho }_{v}-\overline{\rho }_{s}\right) 
\frac{r-R}{w},\;r<R \\ 
\overline{\rho }_{v},\;r>R+w%
\end{array}%
\right.  \\
\chi \left( \mathbf{r}\right) & =\left\{ 
\begin{array}{c}
\chi _{s},\;r<R \\ 
\chi _{s}-\chi _{s}\frac{r-R}{w},\;r<R \\ 
0,\;r>R+w%
\end{array}%
\right.   \notag
\end{align}%
This piecewise-linear model can be extended to give arbitrarily complex
approximations and will play an important role below. For now, it is
substituted into the expression for the free energy which is then simplified
to get%
\begin{eqnarray}
\Omega \left[ \rho \right] -\Omega _{v} &\simeq &\frac{4\pi }{3}R^{3}\left(
\omega _{s}-\omega _{v}\right)   \notag \\
&+&4\pi R^{2}\left( w\overline{\omega }\left( \overline{\rho }_{s},\overline{%
\rho }_{v},\chi _{s}\right) +\frac{1}{2w}K^{\rho \rho }\left( \overline{\rho 
}_{s}-\overline{\rho }_{v}\right) ^{2}+\frac{1}{w}K^{\rho \chi }\chi
_{s}\left( \overline{\rho }_{s}-\overline{\rho }_{v}\right) +\frac{1}{2w}%
K^{\chi \chi }\chi _{s}^{2}\right)   \notag \\
&+&O(Rw,w^{2})
\end{eqnarray}%
with%
\begin{align}
\Omega _{v}& =\int \omega \left( \overline{\rho }_{v},0\right) d\mathbf{r} \\
\overline{\omega }\left( \overline{\rho }_{s},\overline{\rho }_{v},\chi
\right) & =\int_{0}^{1}\omega \left( \overline{\rho }_{s}+\lambda \left( 
\overline{\rho }_{v}-\overline{\rho }_{s}\right) ,\left( 1-\lambda \right)
\chi \right) d\lambda   \notag
\end{align}%
In the present discussion, it is assumed that the clusters are not small (in
the sense that $w/R\ll 1$) so that lower order terms in the expression for $%
\Omega $ can be neglected.

The critical cluster is a stationary point of the free energy so that the
width is determined from 
\begin{equation}
0=\overline{\omega }\left( \overline{\rho }_{s},\overline{\rho }_{v},\chi
_{s}\right) -\frac{1}{2w^{2}}\left( K^{\rho \rho }\left( \overline{\rho }%
_{s}-\overline{\rho }_{v}\right) ^{2}+2K^{\rho \chi }\chi _{s}\left( 
\overline{\rho }_{s}-\overline{\rho }_{v}\right) +K^{\chi \chi }\chi
_{s}^{2}\right) 
\end{equation}%
and the radius from%
\begin{align}
0& =R\left( \omega _{s}-\omega _{v}\right) +2\left( w\overline{\omega }%
\left( \overline{\rho }_{s},\overline{\rho }_{v},\chi _{s}\right) +\frac{1}{%
2w}K^{\rho \rho }\left( \overline{\rho }_{s}-\overline{\rho }_{v}\right)
^{2}+\frac{1}{w}K^{\rho \chi }\chi _{s}\left( \overline{\rho }_{s}-\overline{%
\rho }_{v}\right) +\frac{1}{2w}K^{\chi \chi }\chi _{s}^{2}\right)  \\
& =R\left( \omega _{s}-\omega _{v}\right) +4w\overline{\omega }\left( 
\overline{\rho }_{s},\overline{\rho }_{v},\chi _{s}\right)   \notag
\end{align}%
giving%
\begin{equation}
\Delta \Omega _{vs}\simeq \frac{64\pi }{3\sqrt{2}}\frac{\overline{\omega }%
^{3/2}\left( \overline{\rho }_{s},\overline{\rho }_{v},\chi _{s}\right) }{%
\left( \omega _{v}-\omega _{s}\right) ^{2}}\left( K^{\rho \rho }\left( 
\overline{\rho }_{s}-\overline{\rho }_{v}\right) ^{2}+2K^{\rho \chi }\chi
_{s}\left( \overline{\rho }_{s}-\overline{\rho }_{v}\right) +K^{\chi \chi
}\chi _{s}^{2}\right) ^{3/2}
\end{equation}%
The same model applied to the vapor-liquid and liquid-solid critical
clusters gives%
\begin{align}
\Delta \Omega _{vl}& \simeq \frac{64\pi }{3\sqrt{2}}\frac{\overline{\omega }%
^{3/2}\left( \overline{\rho }_{l},\overline{\rho }_{v},0\right) }{\left(
\omega _{v}-\omega _{l}\right) ^{2}}\left( K^{\rho \rho }\left( \overline{%
\rho }_{l}-\overline{\rho }_{v}\right) ^{2}\right) ^{3/2} \\
\Delta \Omega _{ls}& \simeq \frac{64\pi }{3\sqrt{2}}\frac{\overline{\omega }%
^{3/2}\left( \overline{\rho }_{s},\overline{\rho }_{l},\chi _{s}\right) }{%
\left( \omega _{l}-\omega _{s}\right) ^{2}}\left( K^{\rho \rho }\left( 
\overline{\rho }_{s}-\overline{\rho }_{l}\right) ^{2}+2K^{\rho \chi }\chi
_{s}\left( \overline{\rho }_{s}-\overline{\rho }_{l}\right) +K^{\chi \chi
}\chi _{s}^{2}\right) ^{3/2}  \notag
\end{align}%
Finally, let us make the further approximation that the density of the
liquid and solid states are almost the same so that 
\begin{eqnarray}
\Delta \Omega _{vl} &\simeq &\frac{64\pi }{3\sqrt{2}}\frac{\overline{\omega }%
^{3/2}\left( \overline{\rho }_{s},\overline{\rho }_{v},0\right) }{\left(
\omega _{v}-\omega _{l}\right) ^{2}}\left( K^{\rho \rho }\left( \overline{%
\rho }_{s}-\overline{\rho }_{v}\right) ^{2}\right) ^{3/2} \\
\Delta \Omega _{ls} &\simeq &\frac{64\pi }{3\sqrt{2}}\frac{\overline{\omega }%
^{3/2}\left( \overline{\rho }_{s},\overline{\rho }_{s},\chi _{s}\right) }{%
\left( \omega _{l}-\omega _{s}\right) ^{2}}\left( K^{\chi \chi }\chi
_{s}^{2}\right) ^{3/2}  \notag
\end{eqnarray}%
Double nucleation will be possible if $\omega _{s}<\omega _{l}<\omega _{v}$
and will be the energetically preferred pathway provided $\Delta \Omega
_{vs}>\Delta \Omega _{vl}$ which in turn is more likely if one or more of
these conditions is fulfilled:

\begin{enumerate}
\item The barrier for a direct transition is larger than that for an
indirect transition $\overline{\omega}\left( \overline{\rho}_{s},\overline{%
\rho}_{v},\chi_{s}\right) >\overline{\omega}\left( \overline{\rho}_{l},%
\overline{\rho}_{v},0\right) ,\overline{\omega}\left( \overline{\rho}_{l},%
\overline{\rho}_{v},0\right) $ (It was this condition that was studied
previously by Lutsko and Nicolis\cite{LN}).

\item $\omega_{v}-\omega_{l} >> \omega_{v}-\omega_{s}$

\item $K^{\rho\chi},K^{\chi\chi}$ are not small compared to $K^{\rho \rho}$.
\end{enumerate}

Clearly, the factor $\left( \omega _{v}-\omega _{l}\right) ^{2}$ occurring
in the denominator of $\Delta \Omega _{vl}$ can be important in raising the
vapor-liquid barrier compared to the vapor-solid barrier. On the other hand,
the factor involving the gradient coefficients is always going to be smaller
for the vapor-liquid cluster than for the vapor-solid cluster due to the
terms related to crystallinity.

\section{Gradient coefficients and planar interfaces}

In the standard CNT model, the excess free energy of a cluster is the sum of
a bulk term and a surface term with the latter being proportional to the
planar surface tension at coexistence. In the same way, for the model
considered here, the planar surface tension will play a key role in
determining the gradient coefficients.

\subsection{Gradient coefficients}

Since the second-gradient approximation is the result of a formal expansion
of the free energy, exact expressions for the gradient coefficients exist
(see Appendix \ref{GradCoeff}). Their evaluation requires full knowledge of
the direct correlation function in the bulk system for all densities and
crystallinities which is of course not known. While reasonable models could
be used to make approximate evaluations of the coefficients, the results
would probably be disappointing when used to evaluate physical quantities
such as the liquid-solid or vapor-solid planar surface tension simply
because the squard-gradient model is a crude approximation. In fact, the
relevant interfaces tend to have widths of only a few times the typical
atomic separation so that the assumption of slowly-varying order parameters
that underlies the SGA is unlikely to be very good, although for the
particular case of the liquid-vapor interface which involves only a single
order parameter and no structural change, it is actually rather good\cite{Lutsko_InPrep}. Thus, while it would be interesting to make good evaluations of
the exact expressions for the coefficients, the approach used here is more
pragmatic. First, it is noted that a systematic expansion in crystallinity
and density gives%
\begin{eqnarray}
K^{\rho \rho }\left( \overline{\rho },\chi \right)  &=&K^{\rho \rho }\left(
0,0\right) \left( 1+O\left( \chi ,\overline{\rho }\right) \right)   \notag \\
K^{\rho \chi }\left( \overline{\rho },\chi \right)  &=&C\left( T\right) 
\overline{\rho }\chi \left( 1+O\left( \chi ,\overline{\rho }\right) \right) 
\notag \\
K^{\chi \chi }\left( \overline{\rho },\chi \right)  &=&C\left( T\right) 
\overline{\rho }^{2}\left( 1+O\left( \chi ,\overline{\rho }\right) \right) 
\end{eqnarray}%
with%
\begin{equation*}
C\left( T\right) \equiv \left[ \frac{\partial ^{2}K^{\rho \chi }\left( 
\overline{\rho },\chi \right) }{\partial \overline{\rho }\partial \chi }%
\right] _{\overline{\rho }=\chi =0}=\left[ \frac{\partial ^{2}K^{\chi \chi
}\left( \overline{\rho },\chi \right) }{\partial \overline{\rho }^{2}}\right]
_{\overline{\rho }=\chi =0}
\end{equation*}%
Second, there is evidence that the density-density coefficient can be well
modeled in the liquid, i.e. for $\chi =0$, by a density-independent quantity%
\cite{Lutsko_InPrep}. Thirdly, explicit calculations in the accessible limit of low
density indicate that all three coefficients are relatively insensitive to
the crystallinity (aside from the explicit factors shown above)\ at least up
to crystallinities about half that of the bulk solid. Finally, Laird has noted\cite{Laird_solid_liquid_hs, Davidchack_crystal_melt} that the structural properties tend to be dominated by
the short-range repulsion of the pair potential so that the liquid-solid
surface tension can be approximated by that of the hard-sphere solid
evaluated with an effective hard-sphere diameter giving the useful
approximation that the excess surface free energy for a planar liquid-solid
interface is%
\begin{equation}
\gamma _{ls}\approx 0.617d^{2}\left( T\right) /k_{B}T
\end{equation}%
. This suggests that $K^{\rho \chi }$ and $K^{\chi \chi }$ should be
dominated by hard-sphere contributions which would imply that they scale
linearly with temperature. It is also consistent with the model free energy
functional used here in which \emph{all} of the dependence of the free
energy on the crystallinity enters through the hard-sphere part of the free
energy. All of this suggests a simple approximation for the structural
coefficients along the lines of%
\begin{eqnarray}
K^{\rho \chi }\left( \overline{\rho },\chi \right)  &\approx &C_{\overline{%
\rho }\chi }\overline{\rho }\chi k_{B}T\sigma ^{6} \\
K^{\chi \chi }\left( \overline{\rho },\chi \right)  &\approx &C_{\chi \chi }%
\overline{\rho }^{2}k_{B}T\sigma ^{6}  \notag
\end{eqnarray}%
In a final simplification, the low-density limit gives $C_{\overline{\rho }%
\chi }=C_{\chi \chi }$.

\begin{figure*}[tbp]
\includegraphics[angle=-0,scale=0.4]{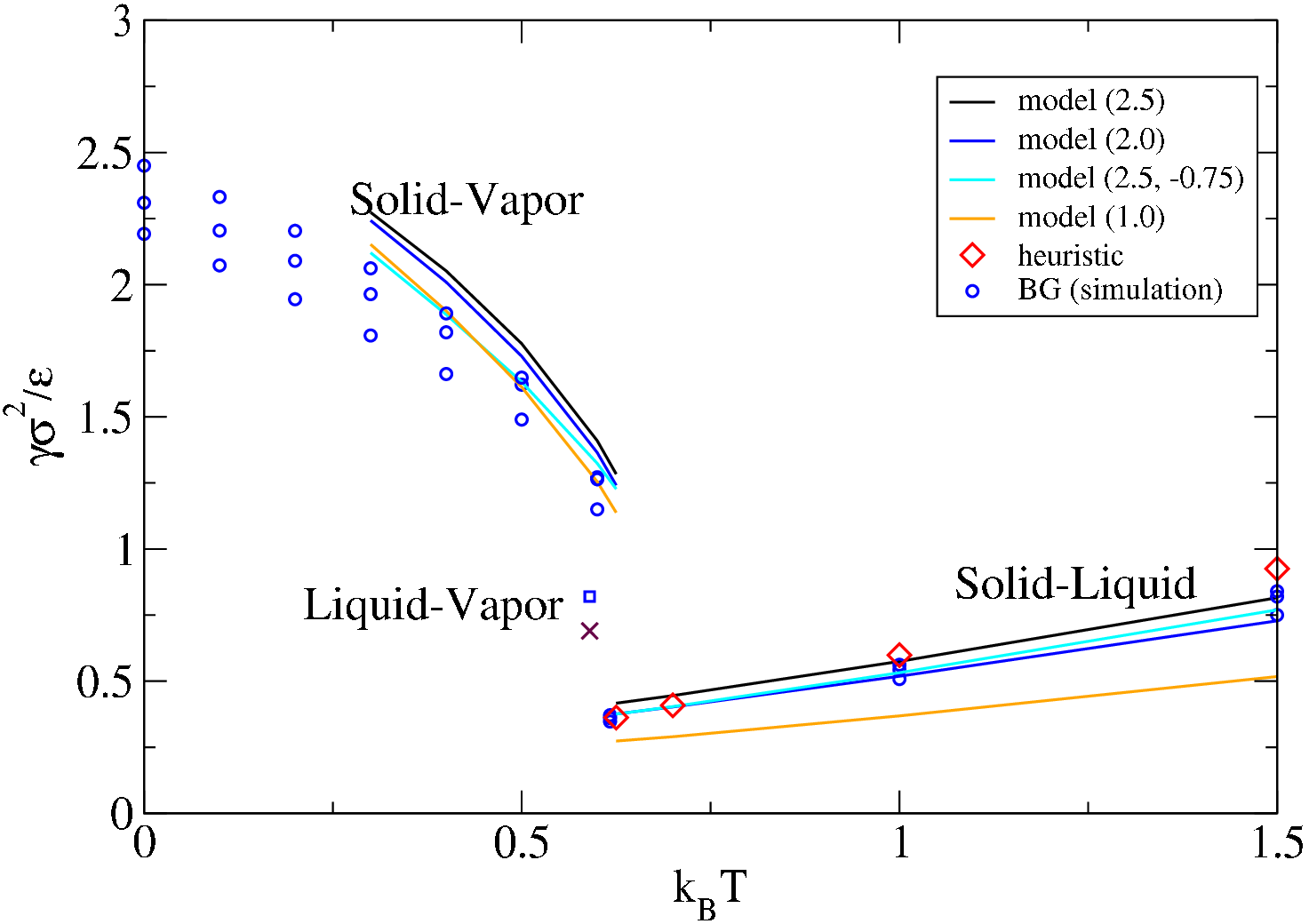} %%{surface_tensions_LJBG}
\caption{Excess surface free energy for the solid-vapor, liquid-vapor and
liquid-solid interfaces for a system interacting via the LJBG pair
potential. The small symbols are simulation data taken from Broughton and
Gilmer\protect\cite{BG4} (solid-vapor and liquid-vapor) and from Davidchack
and Laird\protect\cite{Davidchack_crystal_melt}(solid-liquid). The diamonds
show the Laird approximation for the solid-liquid planar surface tension%
\protect\cite{Laird_solid_liquid_hs}. The simulation results at each
temperature are, from highest to lowest, for the 111, 100 and 110 planes
respectively. }
\label{figLJBG}
\end{figure*}

\subsection{Planar interfaces}

Figure \ref{figLJBG} shows the result of using this model to calculate the
liquid-solid, vapor-solid and vapor-liquid surface tensions at various
temperatures as compared to the available simulation data. It is evident
that, e.g., choosing $C_{\overline{\rho }\chi }=C_{\chi \chi }$ to give,
e.g., a value of the liquid-solid surface tension in agreement with a single
point of either simulation or the Davidchack-Laird approximation is enough
to give a reasonable description of the liquid-solid and vapor-solid surface
tensions over a range of temperatures. Furthermore, reasonable values for
all of the physical quantities are found for a range of choices of the
coefficient so that the model is relatively robust with respect to this
choice.

\begin{figure*}[tbp]
\includegraphics[angle=-0,scale=0.4]{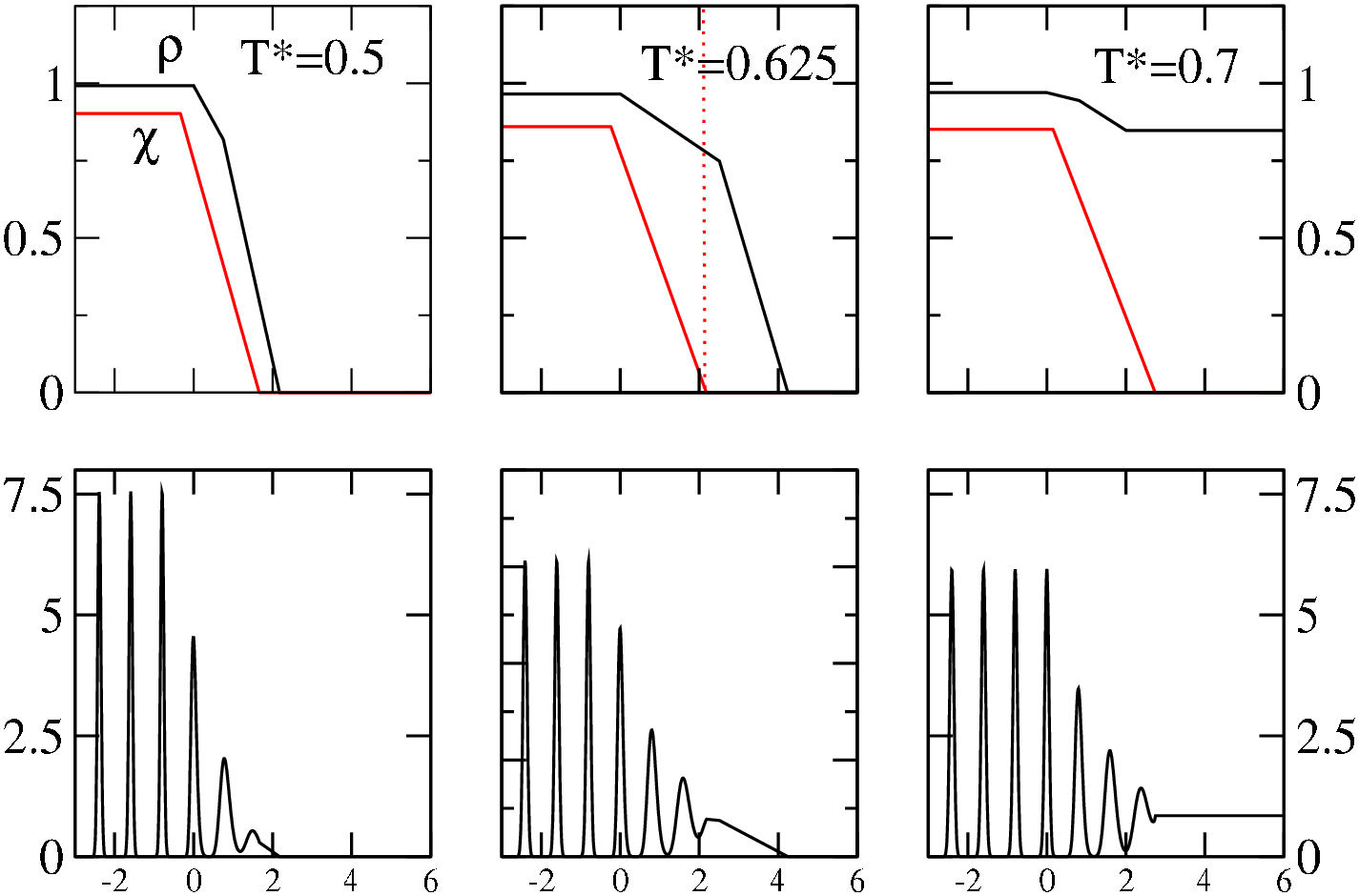} %%{profiles}
\caption{The order parameters (upper panels) and the 100 planar-averaged
density (lower panels) for the solid-fluid interface at three temperatures
which are, respectively, below, near and above the triple point. In the
upper panels, the higher curve is $\overline{\protect\rho}(z)$ and the lower
curve is $\protect\chi(z)$. All quantities are in reduced units.}
\label{figProfiles}
\end{figure*}

Figure \ref{figProfiles} shows the planar profiles calculate using $C_{%
\overline{\rho }\chi }=C_{\chi \chi } = 2$ for a temperature below the
triple point, one near the triple point and one above the triple point. The
profiles above and below the triple point share common features: the width
and position of the transition regions for both the average density and the
crystallinity are roughly the same and the overall width of the interfaces
is about three atomic planes. The profile near the triple point is broader
and is actually composed of two distinct regions: the first part in which
there is a modest drop in density and the crystallinity goes to zero, and
the second region in which the density drops to that of the vapor with the
crystallinity equal to zero. The first region has the nature of a
solid-liquid transition while the second has the structure of a liquid-vapor
transition. In fact, this can be interpreted as a caricature of a wetted
interface. True wetting, with a liquid region of finite width, is not
possible in this model because the long-ranged van der Waals forces that
give rise to the fluid are not being explicitly modeled. Hence, the results
shown are as close as this model can come to representing wetting -
basically, wetting with the bulk fluid region having zero width.

\section{Vapor-Crystal nucleation}

\subsection{General considerations}

Given a model for the free energy of interfacial systems it is now possible
to consider the process of nucleation. In this context, nucleation means the
formation of clusters that eventually become super-critical. As mentioned
above, the most important issue that arises in vapor-solid nucleation is the
description of the nucleation pathway, for which there are at least three
possibilities. The first is the conventional pathway in which the
subcritical cluster has essentially bulk solid properties (density and
crystallinity) except for an interfacial region. When the clusters are
small, they are subcritical and the free energy increases with cluster size
until a transition state - a critical cluster - is reached, after which
further growth lowers the free energy. This pathway is therefore
characterized by a single nucleation barrier and by near-bulk solid
properties of the interior of the cluster. The second pathway involves
double nucleation. First, a purely liquid-like cluster forms and grows until
it reaches a critical size after which it is supercritical and stable with
respect to the vapor. Within this liquid cluster, a second cluster forms,
this time having the properties of the bulk-solid, and goes through a
similar process of subcritical growth reaching a transition state and then
becoming stable with respect to both the vapor and the liquid. This pathway
is therefore characterized by two nucleation events and two energy barriers.
The final possibility is termed transient two-step nucleation and is in some
sense intermediate between the classical and double-nucleation scenarios.
Sufficiently small clusters, which are of course sub-critical, are
liquid-like but at a certain size, below the critical size for the liquid,
the crystallinity begins to increase so that there is a single critical
cluster, perhaps more solid-like than liquid-like, and a single energy
barrier to be crossed on the path towards crystallization. In this case, the
liquid need not even be stable with respect to the vapor - indeed, there is
not even a priori reason why the liquid must exist as a thermodynamic state
(e.g. minimum of the bulk free energy) at all.

The question therefore arises as to how one uses the free energy model to
characterize the nucleation pathway?\ Nucleation is of course a rare,
noise-driven event and a dynamical description would seem most appropriate.
In fact, in this sense, characterizing nucleation pathways is similar to the
problem of characterizing chemical reaction pathways, for which the same
issues occur.

\subsection{Double Nucleation}

The first step in characterizing any structural transition or reaction is
the identification of the saddle points of the free energy surface, i.e. the
transition states. In general, the beginning state (the bulk vapor) and the
end state (the bulk crystal) are local minima of the free energy and the
transition states are the critical nuclei which define the energy barrier
separating the minima. Here, it is assumed that the relevant density
distributions are spherically symmetric and the order parameter fields, $%
\overline{\rho }\left( r\right) $ and $\chi \left( r\right) $ are again
modeled by piecewise continuous functions,%
\begin{equation*}
\overline{\rho }\left( r\right) =\left\{ 
\begin{array}{c}
\rho _{0},\;r<R \\ 
\rho _{0}+\left( \rho _{1}-\rho _{0}\right) \frac{r-R}{w_{0}},\;R<r<R+w_{0}
\\ 
\rho _{1}+\left( \rho _{2}-\rho _{1}\right) \frac{r-R-w_{0}}{w_{1}}%
,\;R+w_{0}<r<R+w_{0}+w_{1} \\ 
...%
\end{array}%
\right.
\end{equation*}%
with a similar model for the crystallinity. In order to refer to different
realizations of these models, the notation $m(i,j)$ will be used indicating
that there are $i$ links for the density profile (which means $1+2i$
parameters since there is the initial radius and then one density and one
width for each link) and $j$ links for the crystallinity profile, giving $%
1+2j$ parameters for a total of $2+2i+2j$ parameters. The free energy is
then a function of those parameters and the transition states are identified
by searching the $2+2i+2j$ parameter space for stationary points of the free
energy surface. This is done using standard eigenvector-following techniques%
\cite{Wales}.

Three specific models will be studied here. First is the "CNT" model in
which there is a single link in both the density and crystallinity profiles
together with the additional constraint that the radii and widths of the two
profiles are the same and that the density and crystallinity in the bulk
region are the same as for the bulk crystal for the given temperature and
chemical potential. This model therefore depends on only two parameters (the
radius and width of the profiles) and is the simplest possible model. The
other models studied will be $m(1,1)$, a single link for each profile, and $%
m(2,1)$, in which there are two links in the density profile. As for the
planar interfaces described above, this is necessary to allow for wetting of
the surface and will be seen to lead to a substantial reduction of the free
energy over the single link model. On the other hand, including additional
links in the crystallinity has very little effect. (For example, at $T=0.6$
and $S=1.25$, the change in free energy of the critical nucleus found using
the $m(2,1)$ model and the $m(2,2)$ model is on the order of one percent.)

\begin{figure*}[tbp]
\includegraphics[angle=-0,scale=0.4]{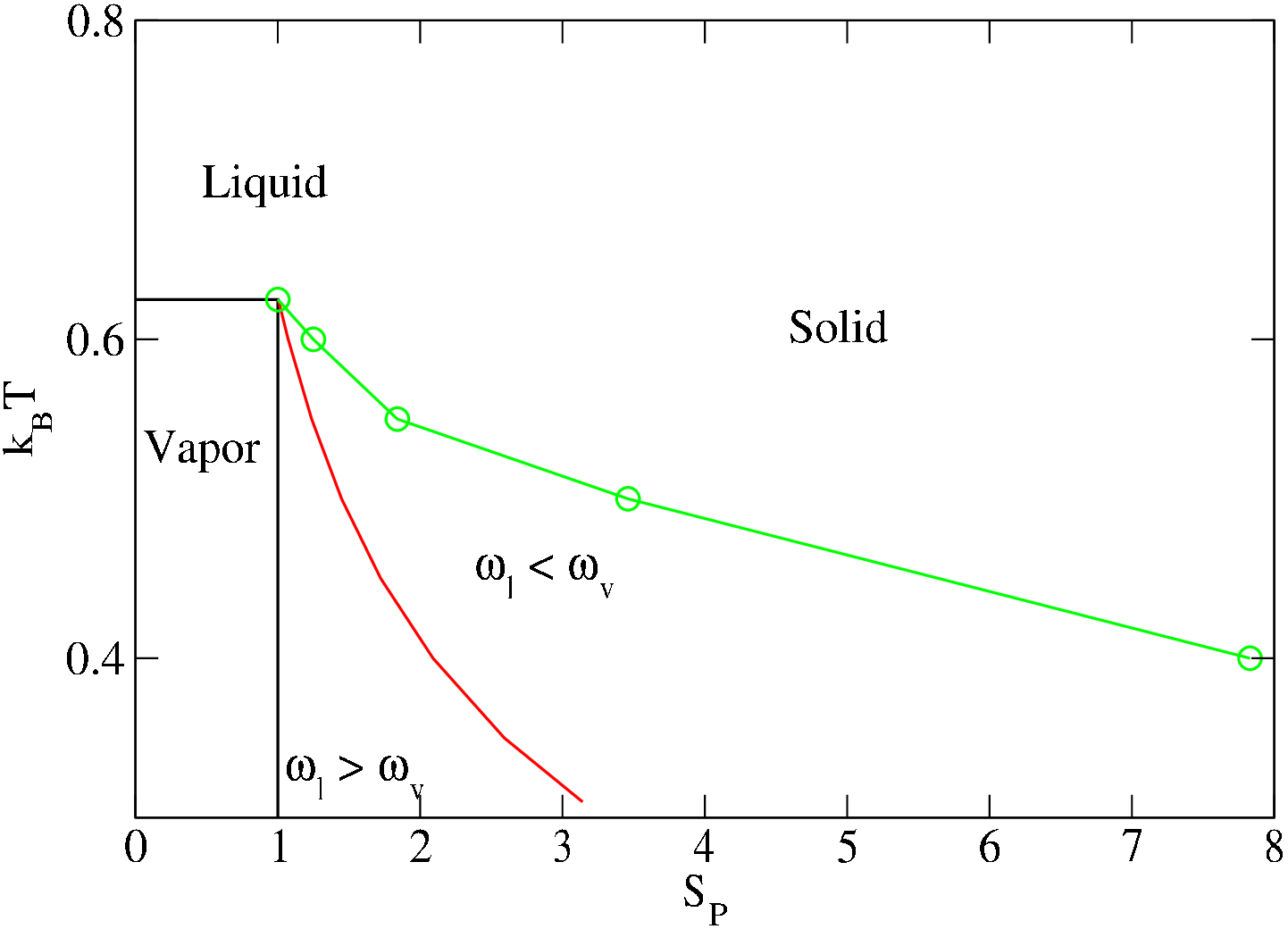} %%{S-kT-LJBG}
\caption{Phase diagram for the LJ potential with (vapor-solid)
supersaturation as the independent variable. Double nucleation occurs in the
region to the right and above the broken line: between the broken line and
red line, the liquid is metastable but double nucleation has a higher energy
barrier than single-step nucleation of the solid. }
\label{supersat}
\end{figure*}

The interesting feature of this problem is that for sufficiently high
supersaturations, both the liquid and the solid are more stable than the
vapor so that there are three transition states:\ one corresponding to the
vapor-liquid transition, another to the vapor-solid transition and a third
to the liquid-solid transition. All of this follows simply from the bulk
free energies as discussed above and illustrated in Fig.\ref{supersat}. The
question addressed here is which of the possible paths will occur:\ direct
transition from vapor to solid or a two-step transition via first a
vapor-liquid transition and then a liquid-solid transition. It is assumed
that whichever transition involves the lowest free energy barriers will
dominate.

Tables \ref{T1} and \ref{T2} give the relevant free energy barriers for
transitions at different supersaturations for $k_{B}T=0.5$ and $0.6$
respectively. It is found that for low supersaturation, corresponding to the
case of large critical nuclei, the barrier for the direct vapor-solid
transition is lower than that for the vapor-liquid transition. However, at
larger supersaturations, the vapor-liquid transition becomes less costly
thus implying that the double-nucleation scenario is favored.

\begin{figure*}[tbp]
\includegraphics[angle=-0,scale=0.4]{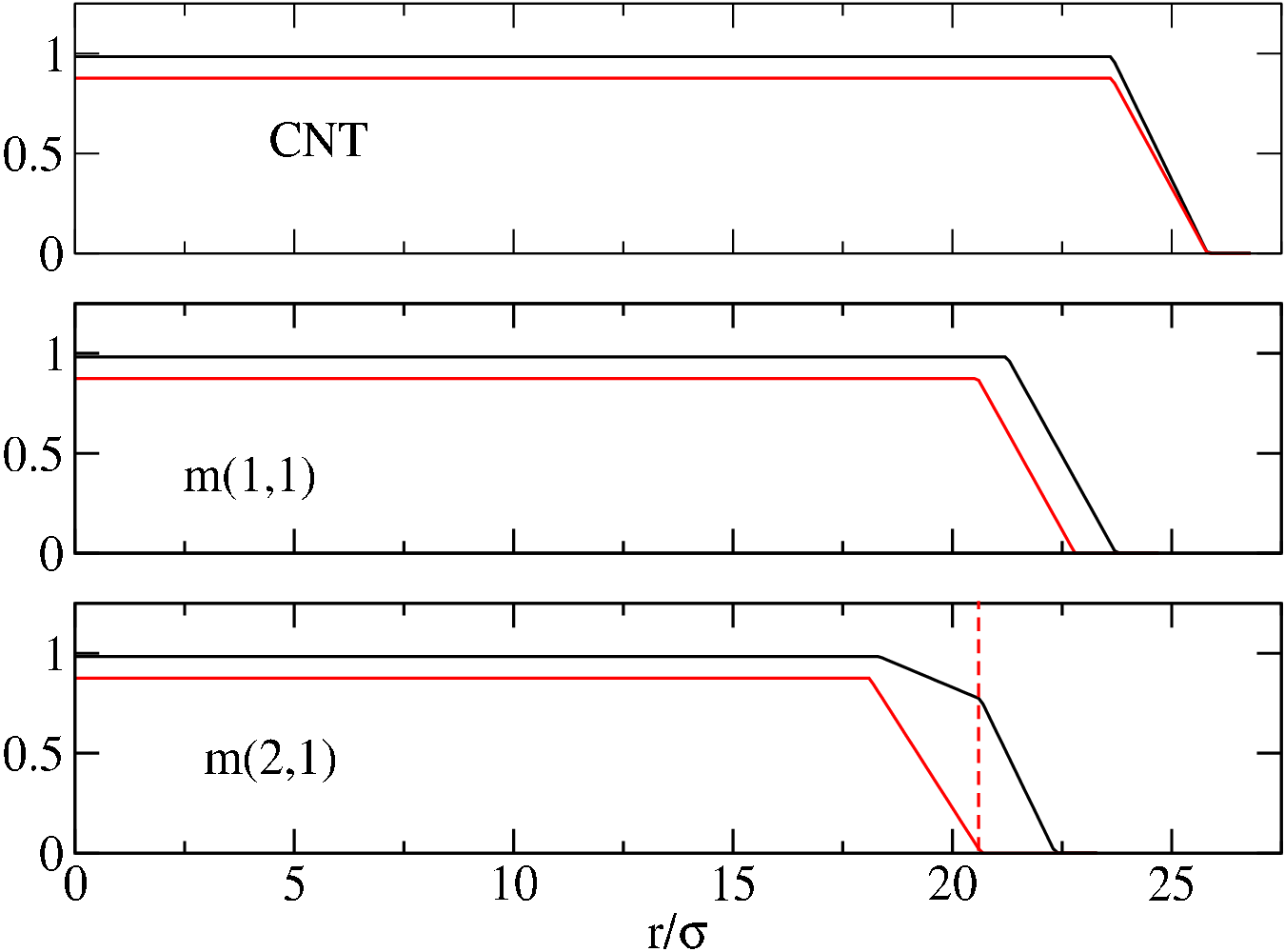} %%{Clusters1}
\caption{The structure of the critical nucleus for $T^*=0.6$ and $S=1.25$ as
determined using the CNT, m(1,1) and m(2,1) parameterizations of the
profiles. In each figure, the upper curve is the average density and the
lower curve is the crystallinity. In the figure for the m(2,1) model, a
dashed line is drawn at the radius where the crystallinity becomes zero: the
fact that the second link of the density profile begins at nearly this point
is a clear illustration of the role of the intermediate liquid state in
wetting the cluster.}
\label{Clusters1}
\end{figure*}

\begin{figure*}[tbp]
\includegraphics[angle=-0,scale=0.4]{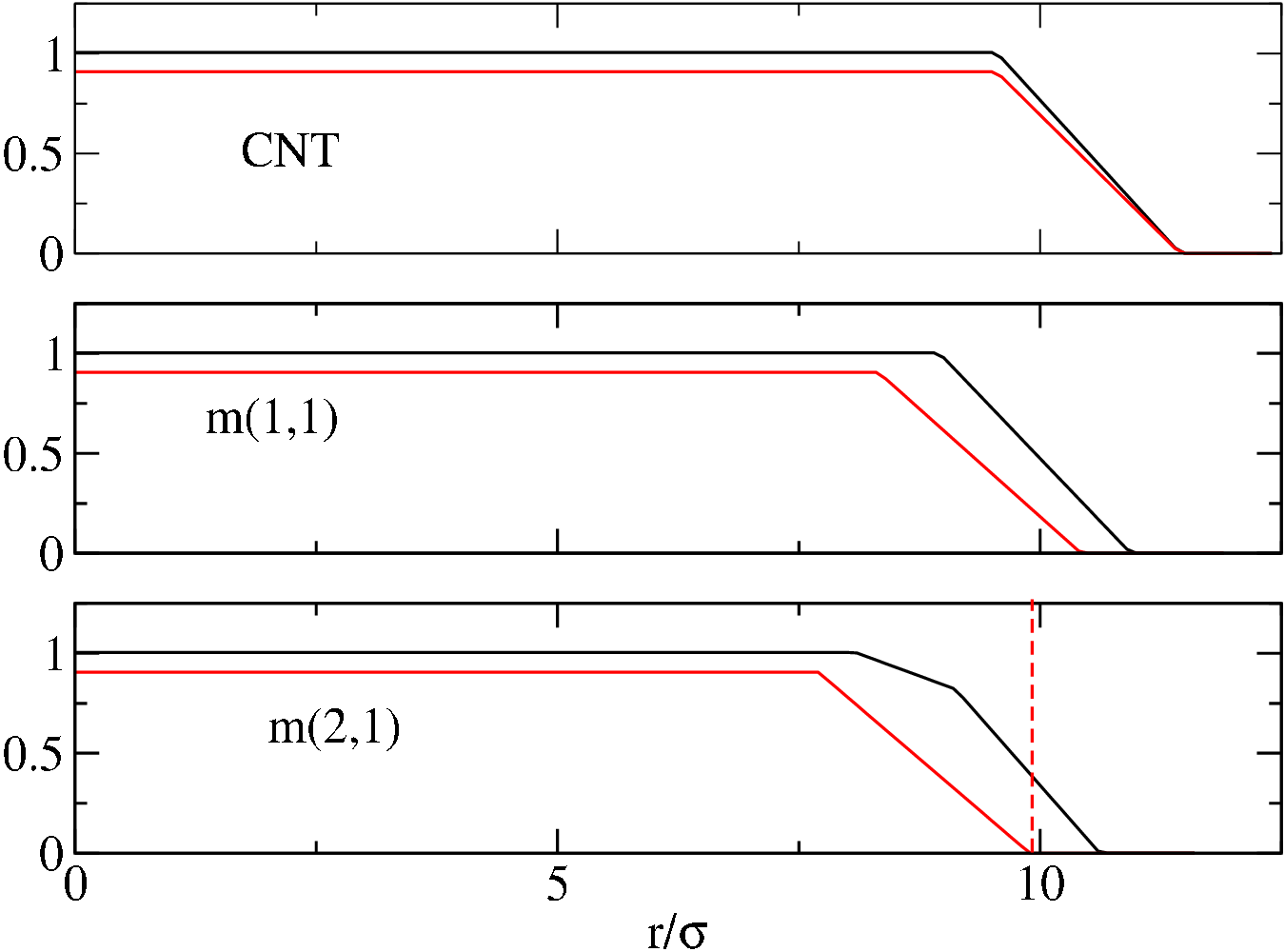} %%{Clusters2}
\caption{The same as Fig. \protect\ref{Clusters1} for $T^*=0.5$ and $S=2.08$%
. Desipite the existence of a metastable intermediate liquid, there is no
significant wetting behaviour in this case.}
\label{Clusters2}
\end{figure*}

Figures \ref{Clusters1} show the structure of the critical nuclei for a few
cases. In all cases, the crystallinity drops to zero before the density
reaches that of the bulk vapor so, in some sense, the crystalline interior
is separated from the bulk vapor by a liquid buffer. This is especially true
if one considers that the structure is essentially liquid-like for
crystallinities less than about 0.3. However, at the higher temperature, a
more distinctive wetting of the cluster is seen whereby the crystallinity
drops to zero over a region in which the density drops from a solid-like to
a liquid-like value followed by a drop of the density from liquid to vapor
values. This is analogous to the planar wetting illustrated above and shows
that this is also a feature of the critical clusters near the triple point.

\bigskip

\begin{table}
\caption{\label{T1}Properties of critical nuclei for the LJBG interaction as
a function of supersaturation at $k_{B}T=0.5$. The supersaturation and
lattice density are given in the first two rows followed by the excess free
energy for the nuclei for vapor-liquid (VL), vapor-solid (VS) and
liquid-solid(LS) nucleation. The last row gives the difference in bulk free
energies for the various transitions. Blank entries occur for cases where it
was not possible to find a transition state.} 
\begin{ruledtabular}
\begin{tabular}{c | ccc | ccc | ccc}
S & \multicolumn{3}{c|}{2.08} & \multicolumn{3}{c|}{3.46} & \multicolumn{3}{c}{
5.92} \\ 
$\rho _{latt}$ & \multicolumn{3}{c|}{1.00} & \multicolumn{3}{c|}{1.005} & 
\multicolumn{3}{c}{1.01} \\ \hline \hline 
\textbf{Model} & $\beta \Omega _{VL}$ & $\beta \Omega _{SV}$ & $\beta \Omega
_{LS}$ & $\beta \Omega _{VL}$ & $\beta \Omega _{SV}$ & $\beta \Omega _{LS}$
& $\beta \Omega _{VL}$ & $\beta \Omega _{VS}$ & $\beta \Omega _{LS}$ \\ \hline
CNT & 2063 & 1738 & 78 & 356 & 587 & 66 & 135 & 269 & 57 \\ 
m(1,1) & 2053 & 1440 & 73 & 355 & 463 & 62 & 133 & * & 53 \\ 
m(2,1) & 1899 & 1304 & 72 & 335 & * & 61 & 128 & * & 53 \\ \hline
$\Delta \beta \Omega _{Bulk}$ & -0.32 & -0.74 & -0.41 & -0.78 & -1.24 & -0.46
& -1.27 & -1.78 & -0.51%
\end{tabular}
\end{ruledtabular}
\end{table}

\begin{table}
\caption{\label{T2} Same as Table \ref{T1} for $k_{B}T=0.6$.} 
\begin{ruledtabular}
\begin{tabular}{c | ccc | ccc | ccc}
S & \multicolumn{3}{c|}{1.26} & \multicolumn{3}{c|}{1.76} & \multicolumn{3}{c}{
3.68} \\ 
$\rho _{latt}$ & \multicolumn{3}{c|}{0.975} & \multicolumn{3}{c|}{0.98} & 
\multicolumn{3}{c}{0.99} \\ \hline \hline 
\textbf{Model} & $\Omega _{LV}$ & $\Omega _{SV}$ & $\Omega _{LS}$ & $\Omega
_{LV}$ & $\Omega _{SV}$ & $\Omega _{LS}$ & $\Omega _{LV}$ & $\Omega _{SV}$ & 
$\Omega _{LS}$ \\ \hline 
CNT & 4045 & 7013 & 758 & 396 & 1142 & 407 & 59 & 189 & 184 \\ 
m(1,1) & 4033 & 5171 & 708 & 392 & 783 & 379 & 58 & * & 171 \\ 
m(2,1) & 3714 & 4176 & 704 & 368 & * & 377 & 57 & * & 170 \\ \hline 
$\beta \Delta \Omega_{Bulk} $ & -0.128 & -0.227 & -0.099 & -0.408 & -0.547 & -0.139
& -1.029 & -1.252 & -0.223%
\end{tabular}
\end{ruledtabular}
\end{table}

\subsection{Transient liquid state}

Even in regions where double nucleation is not possible, the metastable
liquid state could still play a role in nucleation. This becomes a question
of the nucleation pathway which requires much more information than just the
properties of the critical nucleus. A full description of the pathway would
necessarily have to be dynamical in nature accounting for a variety of
kinetic effects including heat and mass transport. It goes without saying
that such a detailed description, while highly desirable, can be expected to
be very difficult to implement.

Alternatively, one might imagine beginning at the transition state which is
a saddle point of the free energy functional and is characterized by a size
(in numbers of atoms), $N_{c}$. It would be natural then to find other
stationary points under the constraint of fixed number of atoms in the
cluster and thus to trace a path from the critical cluster to the bulk
vapor. This is in fact similar to the methods used in simulations\cite{tWF}.
However, even in the case of the liquid-vapor transition this gives
discontinuous paths whereby the structure changes discontinuously for some
value of $N$. For the liquid-vapor transition, this is a transition from a
well-defined cluster with a liquid-like central density to a much larger
structure with a central density slightly larger than the vapor. In the
present case, the transition observed is from a well-defined crystalline
cluster to one with zero crystallinity.

An alternative, widely used in quantum chemistry and in the study of
clusters, is the construction of steepest descent pathways away from the
transition state. These are expected to be approximations to the dynamical
pathway, especially in the case of quasi-equilibrium dynamics due to some
sort of damping:\ an example might be the behavior of colloids (which can
often be modeled as simple fluids) or macromolecules in solution. Many
different methods are used to construct the steepest descent paths (also
commonly called the Minimum Free Energy Pathway or MFEP) including heuristic
methods such as the Nudged Elastic Band and the String Method. Both methods
have been applied to nucleation problems\cite{Lutsko_JCP_2008, LutskoEPL}
but here I expand on recent work which seeks to construct the exact MFEP for
parameterized density profiles\cite{Lutsko_InPrep}.

All methods of constructing steepest descent paths require the notion of a
distance between two points in parameter space since "steepest descent"
literally means the path for which the energy varies most rapidly per unit
distance in parameter space. The problem is that the various parameters used
here - densities, widths, crystallinity - are incommensurate. However, since
they exist only to specify a density profile, a natural solution is to
define a metric in density-space and then to use this to induce a metric in
parameter space. In the study of fluids, the metric was taken to be the
Euclidean distance between two density profiles, 
\begin{equation}
d\left[ \rho ,\rho ^{\prime }\right] ^{2}=\int \left( \rho \left( \mathbf{r}%
\right) -\rho ^{\prime }\left( \mathbf{r}\right) \right) ^{2}d\mathbf{r}
\end{equation}%
It would be natural to continue to use this definition however, it is
unsuitable for two reasons. The first is simply that it is technically
difficult to evaluate. The second, more importantly, is that it leads to the
metric being sensitive to details of the lattice structure. For example, as
the radius parameter varies there is little variation in the metric until
the radius crosses an atomic shell at which point there is very rapid
variation. This is unacceptable in the present context since the free energy
surface constructed above is based on a separation of length scales
according to which the order parameters vary slowly over atomic length
scales.

In order to motivate a simple alternative more in keeping with the present
approach, note that for the liquid-vapor transition the crystallinity is
identically zero, $\chi \left( \mathbf{r}\right) =0$, so that the Euclidean
distance function becomes%
\begin{equation}
d\left[ \rho ,\rho ^{\prime }\right] ^{2}=\int \left( \overline{\rho }\left( 
\mathbf{r}\right) -\overline{\rho }^{\prime }\left( \mathbf{r}\right)
\right) ^{2}d\mathbf{r}
\end{equation}%
which is the Euclidean distance between the $K=0$ amplitudes in the
expansion of the density. In fact, the parameterization of the density used
here is%
\begin{equation}
\rho \left( \mathbf{r}\right) =\overline{\rho }\left( \mathbf{r}\right)
+\sum_{j\epsilon NN}\overline{\rho }\left( \mathbf{r}\right) \chi \left( 
\mathbf{r}\right) e^{i\mathbf{K}_{j}\cdot \mathbf{r}}+...
\end{equation}%
where the first sum is over the first (nearest-neighbor) shell in
wave-vector space. Hence, the quantity $\overline{\rho }\left( \mathbf{r}%
\right) \chi \left( \mathbf{r}\right) $ is the spatially-varying amplitude
of the smallest non-zero wavevector. It therefore seems reasonable to treat
this on a par with the amplitude of the zero-wavevector component and to
define a distance function as%
\begin{equation}
d\left[ \rho ,\rho ^{\prime }\right] ^{2}=\int \left( \overline{\rho }\left( 
\mathbf{r}\right) -\overline{\rho }^{\prime }\left( \mathbf{r}\right)
\right) ^{2}d\mathbf{r+}\int \left( \overline{\rho }\left( \mathbf{r}\right)
\chi \left( \mathbf{r}\right) -\overline{\rho }^{\prime }\left( \mathbf{r}%
\right) \chi ^{\prime }\left( \mathbf{r}\right) \right) ^{2}d\mathbf{r}
\end{equation}%
which is what will be used henceforth. When the order parameters are
expressed in terms of a collection of scalar parameters as $\overline{\rho }%
\left( \mathbf{r}\right) =\overline{\rho }\left( \mathbf{r;}\Gamma \right) $
and $\chi \left( \mathbf{r}\right) =\chi \left( \mathbf{r;}\Gamma \right) $
this then defines a distance between points in parameter space%
\begin{equation}
d\left[ \Gamma ,\Gamma ^{\prime }\right] ^{2}=\int \left( \overline{\rho }%
\left( \mathbf{r;}\Gamma \right) -\overline{\rho }\left( \mathbf{r;}\Gamma
^{\prime }\right) \right) ^{2}d\mathbf{r+}\int \left( \overline{\rho }\left( 
\mathbf{r;}\Gamma \right) \chi \left( \mathbf{r;}\Gamma \right) -\overline{%
\rho }\left( \mathbf{r;}\Gamma ^{\prime }\right) \chi \left( \mathbf{r;}%
\Gamma ^{\prime }\right) \right) ^{2}d\mathbf{r}
\end{equation}%
Assuming this function is sufficiently continuous, the distance function
implies a metric%
\begin{equation}
g_{ij}\left( \Gamma \right) =\int \left( \frac{\partial \overline{\rho }%
\left( \mathbf{r;}\Gamma \right) }{\partial \Gamma _{i}}\frac{\partial 
\overline{\rho }\left( \mathbf{r;}\Gamma \right) }{\partial \Gamma _{j}}%
\frac{\partial \overline{\rho }\left( \mathbf{r;}\Gamma \right) \chi \left( 
\mathbf{r;}\Gamma \right) }{\partial \Gamma _{i}}\frac{\partial \overline{%
\rho }\left( \mathbf{r;}\Gamma \right) \chi \left( \mathbf{r;}\Gamma \right) 
}{\partial \Gamma _{j}}\right) d\mathbf{r}
\end{equation}%
The steepest descent paths are then determined by 
\begin{equation}
\frac{d\Gamma _{i}}{ds}=-\frac{g_{ij}\left( \Gamma \right)\frac{\partial
\beta \Omega }{\partial \Gamma _{j}}}{\frac{\partial \beta \Omega }{\partial
\Gamma _{a}}g_{ab}\left( \Gamma \right)\frac{\partial \beta \Omega }{%
\partial \Gamma _{b}}}
\end{equation}
This equation is obviously similar to a dynamics consisting of simple
relaxation driven by free energy gradients as discussed in Appendix \ref%
{Dynamics}. However, since nucleation is, fundamentally, a noise-driven
process, there is no reason to expect a priori that the correct path can be
determined by running a deterministic dynamics backwards from the transition
state. The idea of steepest descent is that it includes the idea of being a
most probable path since it is in some sense the most efficient path over
the barrier. (An analogy would be a multidimensional random walker in a
potential field. In order for the walker to pass over a barrier, it must
take some number of improbable steps in the right direction until it reaches
the top of the barrier. The steepest descent path is the one involving the
fewest number of steps. Other paths must cross the same barrier, but by
including more steps, say in the ``wrong'' direction, the probability of
falling backwards towards the local minimum increases.)

\begin{figure*}[tbp]
\includegraphics[angle=-0,scale=0.4]{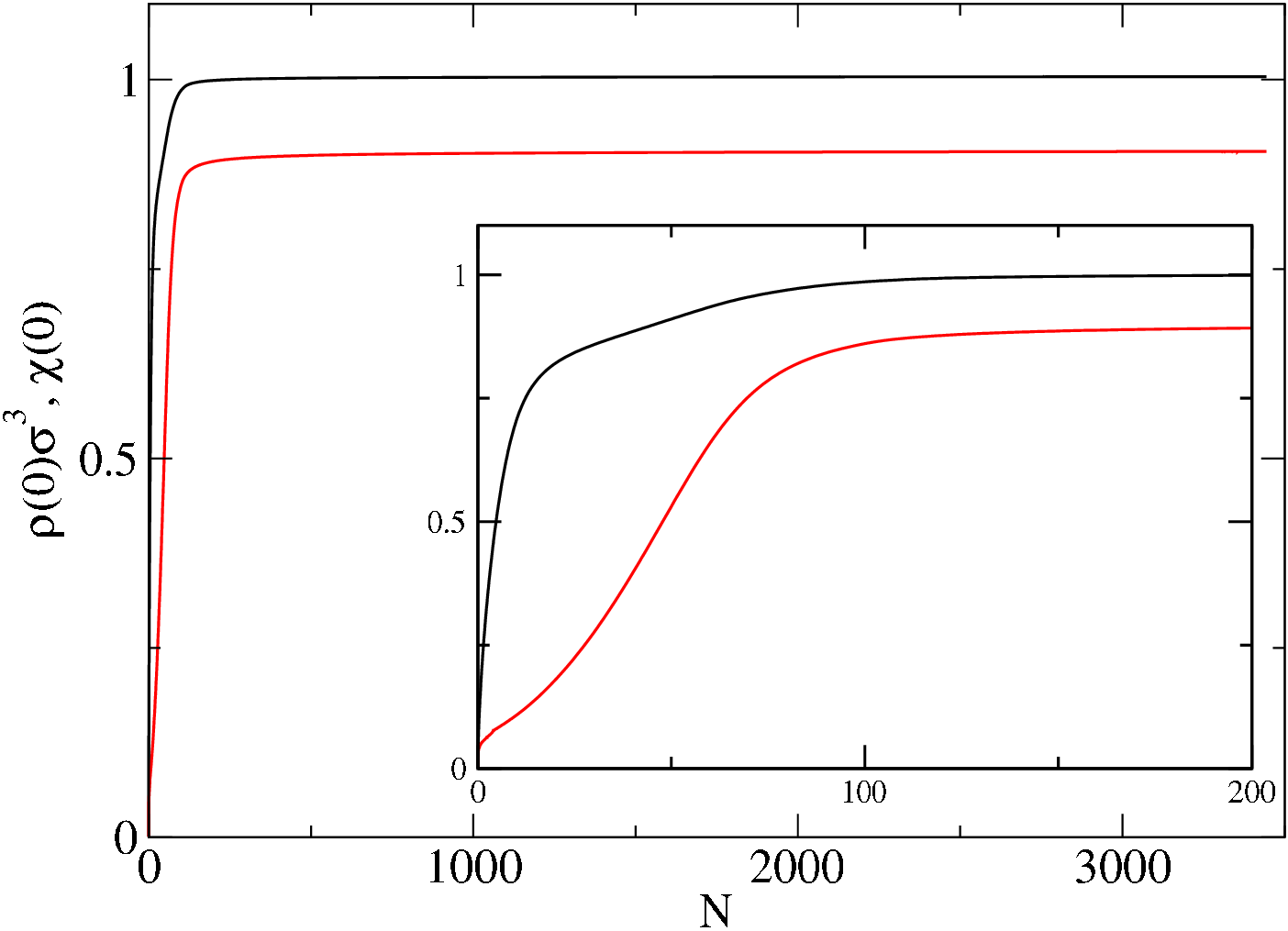} %%{pathT05}
\caption{The average density (upper curve) and crystallinity (lower curve)
at the center of the cluster as a function of cluster size for $T^*=0.5$.
The inset shows an expanded view of the early stages of nucleation.}
\label{path1}
\end{figure*}

\begin{figure*}[tbp]
\includegraphics[angle=-0,scale=0.4]{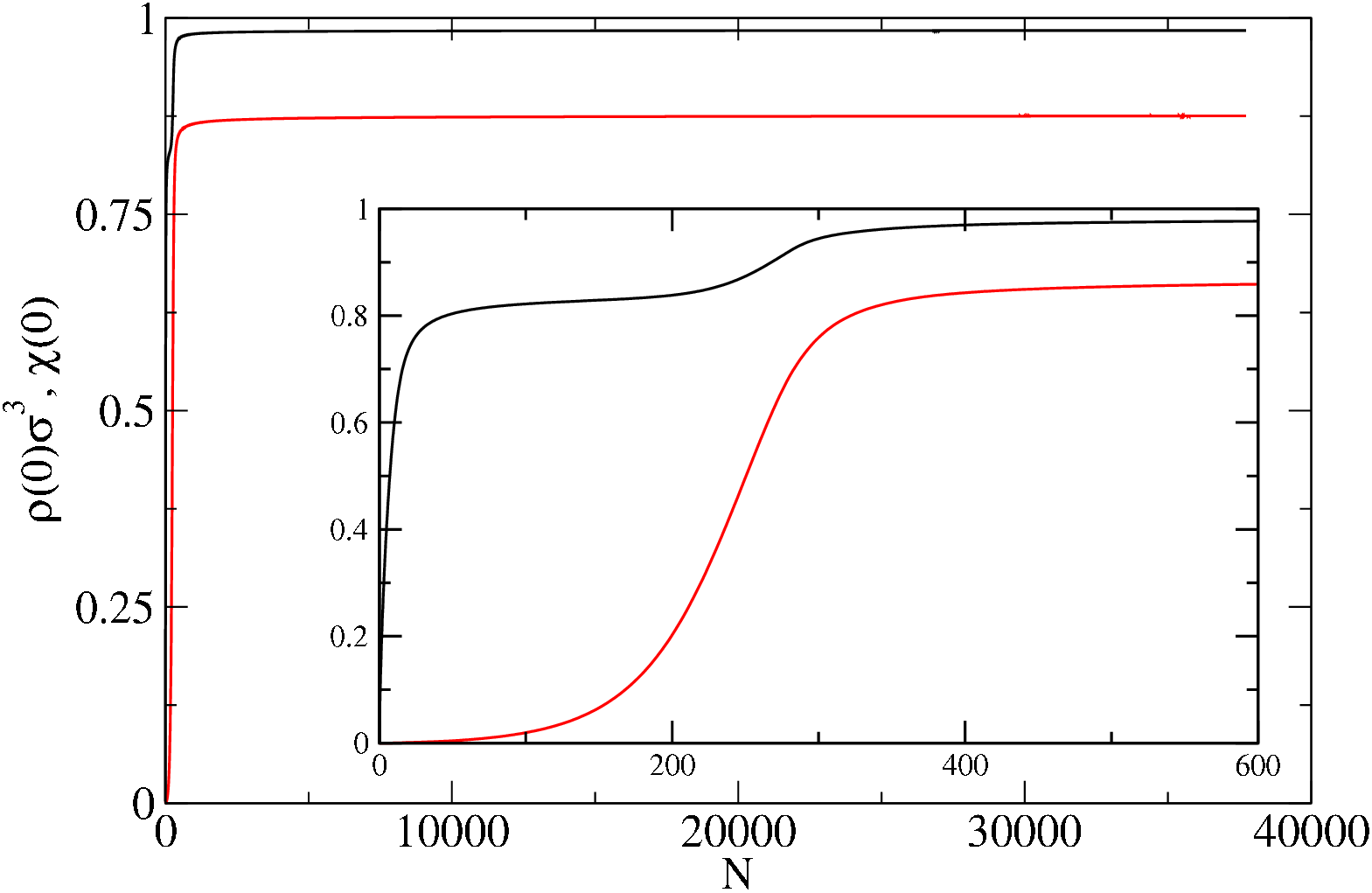} %%{pathT06}
\caption{Same as Fig. \protect\ref{path1} for $T*=0.6$. }
\label{path2}
\end{figure*}

\begin{figure*}[tbp]
\includegraphics[angle=-0,scale=0.4]{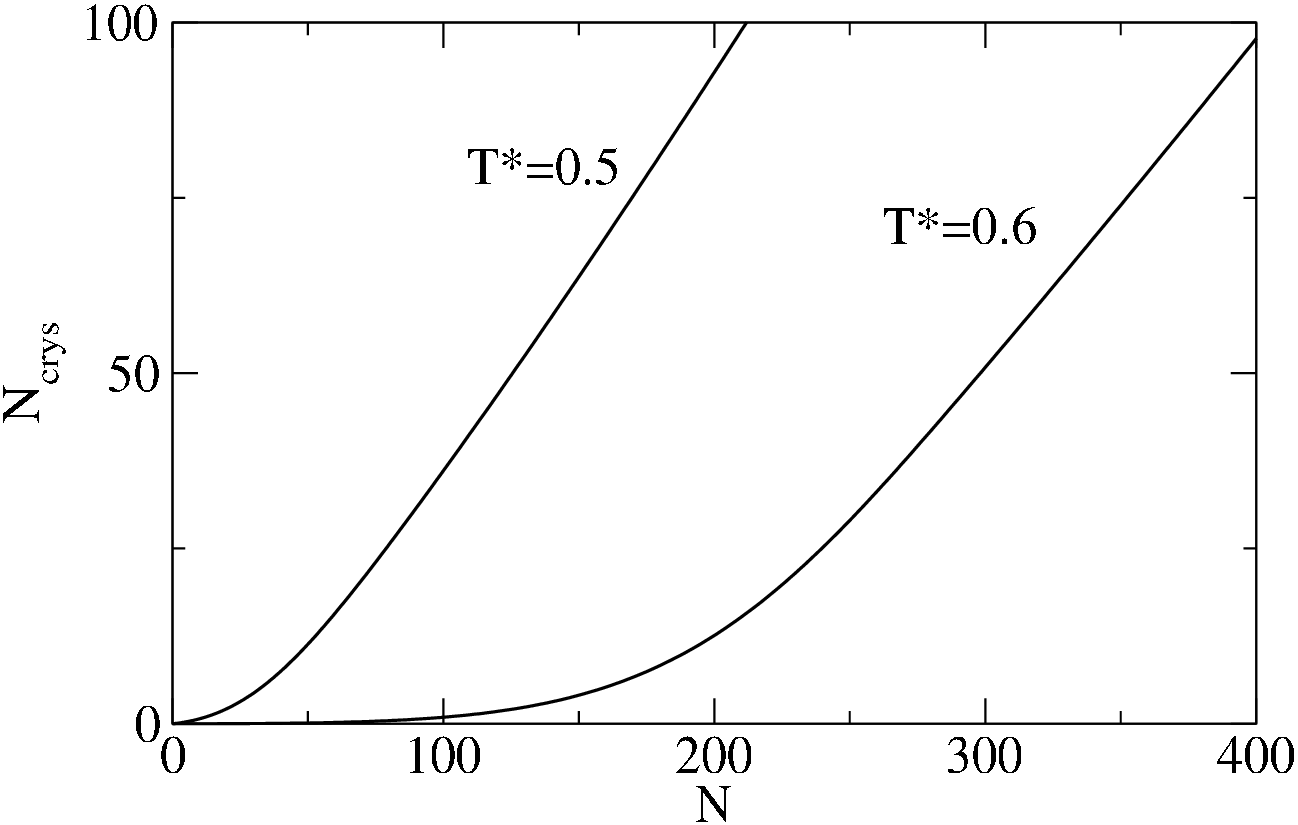} %%{NNcryst}
\caption{The number of crystalline atoms as a function of the total number
of atoms in the cluster showing transient two-step nucleation near the
triple point, $T*=0.6$, and almost completely one-step nucleation away from
the triple point.}
\label{NN}
\end{figure*}

The nucleation pathways have been calculated using this model for values of
the supersaturation such that the metastable liquid is more stable than the
vapor, but below the double-nucleation threshold. Figure \ref{path1} shows
the pathway for $k_{B}T=0.5$ and $S=2.08$ plotted as a function of the the
total number of atoms in the cluster. It is clear that both the central
density and central crystallinity begin at the values of the bulk vapor.
This is because the surface tension depends on the gradient of these
quantities and for very small clusters, the dominance of the surface tension
term over the bulk free energy contribution forces the gradients to be zero.
As the cluster grows, the density increases very rapidly while the
crystallinity increases more slowly. The core of the cluster therefore
densifies more quickly than it crystallizes but the effect is minor. Figure %
\ref{path2} shows the same quantities for $k_{B}T=0.6$, close to the triple
point, and for $S=1.25$. In this case, transient two-step nucleation is
clearly in evidence: the density increases while the crystallinity remains
almost at zero until the cluster consists of well over 100 atoms. These
contrasting behaviors are compared in Fig. \ref{NN} where the number of
\textquotedblleft crystalline\textquotedblright\ atoms, defined as 
\begin{equation}
N_{crys}=\frac{1}{\chi _{bulk}}\int \overline{\rho }\left( \mathbf{r}\right)
\chi \left( \mathbf{r}\right) d\mathbf{r}
\end{equation}%
is plotted as a function of $N$. Near the triple point, the number of
crystalline atoms does not increase appreciably until the cluster is over $%
100$ atoms in size whereas it increases almost immediately at the lower
temperature. This behavior is, incidentally, quite similar to the
observations of ten Wolde and Frenkel who noted that for their model of
globular proteins, the favored nucleation path involved two-step nucleation
near the triple point but that this was not seen further below the triple
point\cite{tWF}.

\section{Conclusions}

The primary goal of this work has been they study of different pathways for
the homogeneous nucleation of crystals from solution based on mean-field,
DFT models. Simply from the bulk thermodynamics, it is possible to divide
the phase diagram into regions for which double nucleation is and is not
possible. When the bulk models are extended to inhomogeneous, multiphase
systems - here via the squared-gradient approximation - it becomes possible
to determine when double nucleation is more energetically favorable than
single-step nucleation. Finally, by studying steepest-descent pathways
connecting the transition states to the bulk states it was possible to
illustrate transient two-step nucleation for the Lennard-Jones fluid. A
summary of these investigations would be that:

\begin{enumerate}
\item there is a lower supersaturation limit for double nucleation: at the
triple point, the limit goes to one and it increases rapidly as the
temperature is lowered.

\item transient two-step nucleation seems to be closely tied to wetting and
to therefore occur most distinctly near the triple point.
\end{enumerate}

Almost all phases of this study could be improved. The underlying DFT model
is highly simplified and in particular the inclusion of a long-ranged van
der Waals term would give a more realistic description of wetting including
a finite wetting-layer thickness. The modeling of the gradient coefficients
is crude and quite empirical. One can estimate them directly from the DFT
model\cite{Lowen1, Lowen2, LutskoGL, OLW1} and this might give a more
realistic dependence on density and crystallinity. The piecewise continuous
models could be used with more than the minimal number of links used here or
some other, perhaps more physical, basis functions could be chosen (or the
Euler-Lagrange equations could be solved directly without parameterizing the
fields). Finally, instead of the Minimum Free Energy Pathways studied here,
a more physically meaningful approach would be to search for the most \emph{%
probable} pathways which requires the introduction of dynamics and noise,
but which seems quite feasible based on the approach of Heymann and
Vanden-Eijnden\cite{MLP}.

\begin{acknowledgments}
I am grateful to Gregoire Nicolis for reading an early draft of this manuscript and for a number of useful suggestions.
This work was supported in part by the European Space Agency under contract
number ESA AO-2004-070.
\end{acknowledgments}

\appendix{}

\section{Bulk solid properties}

\label{Bulk} The calculation of bulk thermodynamic properties for the
hard-sphere solid using the given parameterization is complex. The reason is
that the hard-sphere free energy functional diverges for $\overline{\rho}$
slightly greater than $\rho_{latt}$ - at typical solid densities, the
divergence occurs for $\overline{\rho}-\rho_{latt}\sim10^{-10}$. It might
seem that this is no problem as one simply takes $\overline{\rho}%
=\rho_{latt} $ however, upon closer inspection, this doesn't work.

To explain the problem, let us consider the "correct" way to fix these two
densities. In principle, an equilibrium state must satisfy three conditions:%
\begin{align}
\left. \frac{\partial F\left[ \rho_{\Gamma}\right] }{\partial\overline {\rho}%
}\right\vert _{\rho_{latt},\alpha} & =\mu \\
\left. \frac{\partial F\left[ \rho_{\Gamma}\right] }{\partial\rho_{latt}}%
\right\vert _{\overline{\rho},\alpha} & =0  \notag \\
\left. \frac{\partial F\left[ \rho_{\Gamma}\right] }{\partial\alpha }%
\right\vert _{\rho_{latt},\overline{\rho}} & =0  \notag
\end{align}
Let the solution to these equations be $\overline{\rho}^{\ast},\rho
_{latt}^{\ast},\alpha^{\ast}$. Solution of these three simultaneous
equations is delicate due to the fact that (a)\ $\left\vert \overline{\rho}%
-\rho _{latt}\right\vert \ll1$ and (b) the surprising fact that%
\begin{equation}
\left\vert \left. \frac{\partial F\left[ \rho_{\Gamma}\right] }{\partial%
\overline{\rho}}\right\vert _{\overline{\rho}=\rho_{latt}^{\ast},%
\rho_{latt}^{\ast},\alpha^{\ast}}-\left. \frac{\partial F\left[ \rho_{\Gamma}%
\right] }{\partial\overline{\rho}}\right\vert _{\overline{\rho }%
^{\ast},\rho_{latt}^{\ast},\alpha^{\ast}}\right\vert =O\left( \mu\right)
\end{equation}
In words, the free energy is a \emph{very} rapidly varying function of $%
\overline{\rho}$ for$\overline{\rho}$ near $\rho_{latt}$.

This situation is somewhat better than it seems since in practical
calculations, we are almost always performing some sort of search over, or
tabulation in terms of, the chemical potential. Hence, rather than being
given $\mu$ and having to solve three simultaneous equations, we can often
take, e.g., $\rho_{latt}$ as the independent parameter and search for $%
\overline {\rho}^{\ast},\alpha^{\ast}$ satisfying 
\begin{align}
\left. \frac{\partial F\left[ \rho_{\Gamma}\right] }{\partial\rho_{latt}}%
\right\vert _{\overline{\rho}^{\ast},\rho_{latt},\alpha^{\ast}} & =0
\label{a2} \\
\left. \frac{\partial F\left[ \rho_{\Gamma}\right] }{\partial\alpha }%
\right\vert _{\overline{\rho}^{\ast},\rho_{latt},\alpha^{\ast}} & =0  \notag
\end{align}
We then calculate $\mu$ given these values. (Note however that even
evaluating the various derivatives numerically is quite delicate when the
range over which $\overline{\rho}d^{3}$ can be varied symmetrically might be 
$10^{-10}$ or smaller.)

There is a heuristic which alleviates many of these problems. Intuitively,
one tends to think in terms of \textit{the} density and not to distinguish
between the lattice density and the average density. In fact, in most
calculations for the hard-sphere solid, the approximation $\overline{\rho }%
=\rho _{latt}$ is made and the results appear quite reasonable. Since%
\begin{equation}
\frac{df\left( z,z\right) }{dz}=\left. \frac{\partial f\left( x,y\right) }{%
\partial x}\right\vert _{x=y=z}+\left. \frac{\partial f\left( x,y\right) }{%
\partial y}\right\vert _{x=y=z}
\end{equation}%
this suggests that it must be the case that 
\begin{equation}
\left. \frac{\partial F\left[ \rho _{\Gamma }\right] }{\partial \overline{%
\rho }}\right\vert _{\overline{\rho }=\rho _{latt}^{\ast },\rho
_{latt}^{\ast },\alpha ^{\ast }}+\left. \frac{\partial F\left[ \rho _{\Gamma
}\right] }{\partial \rho _{latt}}\right\vert _{\overline{\rho }=\rho
_{latt}^{\ast },\rho _{latt}^{\ast },\alpha ^{\ast }}\simeq \left. \frac{%
\partial F\left[ \rho _{\Gamma }\right] }{\partial \overline{\rho }}%
\right\vert _{\overline{\rho }^{\ast },\rho _{latt}^{\ast },\alpha ^{\ast
}}=\mu   \label{a1}
\end{equation}%
In fact, this suspicion is borne out in practice. Combining these two
points, and noting that the value of $\alpha $ that makes the free energy
stationary varies slowly as a function of the various densities, an
efficient practical procedure is to choose a value of $\rho _{latt}$, to set 
$\overline{\rho }=\rho _{latt}$, to determine the stationary value of $%
\alpha $ and to use the left hand side of Eq.\ref{a1} to evaluate $\mu $.
This involves a simple, controlled minimization and no need to evaluate the
free energy near the divergence. A proof that Eq. \ref{a1} is exact, or at
least a good approximation, is to my knowledge missing and would be useful.

Finally, an important point is that these technicalities play no role in the
calculations presented here. Since we deal here with interfacial problems,
the important thing is how the free energy varies as the average density
varies from that of the solid down to a relatively low value (that of a
liquid or vapor). Furthermore, in clusters, the interior density is always
somewhat below that of the bulk solid (dwarfing the comparatively tiny
difference between$\overline{\rho }^{\ast }$ and $\rho _{latt}^{\ast }$) .
The only way these technical points would be important would be if we tried
to do a free minimization of the free energy for a planar interface in which
case the algorithm would have to find the correct values for the bulk
system. This would essentially mean solving Eq. \ref{a2} which, because of
the divergences very near the correct solution, leads to numerical
challenges. However, using the piecewise-continuous approximation, we always
set the bulk values "by hand" and thereby avoid this problem

\section{Formal results for gradient coefficients}

\label{GradCoeff} They are given by%
\begin{equation*}
\beta K_{ab}^{\Gamma _{a}\Gamma _{b}}\left( \Gamma \right) =-\frac{1}{2V}%
\int \mathbf{r}_{12,a}\mathbf{r}_{12,b}c\left( \mathbf{r}_{1},\mathbf{r}%
_{2};\Gamma \right) \frac{\partial \rho \left( \mathbf{r}_{1};\Gamma \right) 
}{\partial \Gamma _{a}}\frac{\partial \rho \left( \mathbf{r}_{2};\Gamma
\right) }{\partial \Gamma _{b}}d\mathbf{r}_{1}d\mathbf{r}_{2}
\end{equation*}%
where $c\left( \mathbf{r}_{1},\mathbf{r}_{2};\Gamma \right) $ is the direct
correlation function for the bulk solid. Unfortunately, the DFT models used
here do not give realistic expressions for this quantity although some
useful results are possible. In particular, expanding in the crystallinity
gives%
\begin{eqnarray}
K^{\rho \rho } &=&I_{1}\left( \overline{\rho }\right) +O\left( \chi \right)
\\
K^{\rho \chi } &=&\overline{\rho }\chi I_{2}\left( \overline{\rho }\right)
+O\left( \chi ^{2}\right)  \notag \\
K^{\chi \chi } &=&\overline{\rho }^{2}I_{2}\left( \overline{\rho }\right)
+O\left( \chi \right)  \notag
\end{eqnarray}%
with%
\begin{eqnarray}
I_{1}\left( \overline{\rho }\right) &=&\frac{2\pi }{3}\int_{0}^{\infty
}c\left( r;\overline{\rho }\right) r^{4}dr \\
I_{2}\left( \overline{\rho }\right) &=&\frac{2\pi }{3}N_{1}\int_{0}^{\infty }%
\frac{\sin \left( K_{1}r\right) }{K_{1}}c\left( r;\overline{\rho }\right)
r^{3}dr  \notag
\end{eqnarray}%
which only requires the DCF in the bulk fluid (albeit, at solid densities).
Similarly, expanding in density gives%
\begin{equation}
\beta K_{ab}^{\Gamma _{a}\Gamma _{b}}\left( \Gamma \right) =-\frac{1}{2V}%
\int \mathbf{r}_{12,a}\mathbf{r}_{12,b}\left( 1-e^{-\beta v\left( r\right)
}\right) \frac{\partial \rho \left( \mathbf{r}_{1};\Gamma \right) }{\partial
\Gamma _{a}}\frac{\partial \rho \left( \mathbf{r}_{2};\Gamma \right) }{%
\partial \Gamma _{b}}d\mathbf{r}_{1}d\mathbf{r}_{2}+...
\end{equation}%
and combining the two expansions gives%
\begin{eqnarray}
K^{\rho \rho } &=&I_{1}\left( 0\right) \left( 1+O\left( \chi ,\overline{\rho 
}\right) \right) \\
K^{\rho \chi } &=&\overline{\rho }\chi I_{2}\left( 0\right) \left( 1+O\left(
\chi ,\overline{\rho }\right) \right)  \notag \\
K^{\chi \chi } &=&\overline{\rho }^{2}I_{2}\left( 0\right) \left( 1+O\left(
\chi ,\overline{\rho }\right) \right)  \notag
\end{eqnarray}

\bigskip

and even if they did, the coefficients evaluated from these expressions
might well give poor values of the surface tension due to the truncation of
the gradient expansion.

\bigskip

Coexistence occurs at a given temperature for a single value of the chemical
potential corresponding, by definition, to supersaturation equal to $1$.

\section{Steepest Descent and Dynamics}

\label{Dynamics} Typical dynamical models depend on a distinction between
order parameters which are densities of conserved quantities (such as the
total mass in the canonical ensemble) and those which are densities of
non-conserved quantities (such as the mass in the grand canonical ensemble).
We begin with the latter case of a non-conserved order parameter. Then, the
evolution is often assumed to be of the form 
\begin{equation*}
\frac{d\psi _{t}\left( \mathbf{r}\right) }{dt}=-\Gamma \frac{\delta \Omega %
\left[ \psi _{t}\right] }{\delta \psi _{t}\left( \mathbf{r}\right) }
\end{equation*}%
where $\Gamma $ is a transport coefficient. If space is discretized and we
denote $\psi _{t}\left( \mathbf{r}_{i}\right) =\psi _{ti}$, this takes the
form%
\begin{equation*}
\frac{d\psi _{ti}}{dt}=-\Gamma \frac{\partial \Omega \left[ \psi _{t}\right] 
}{\partial \psi _{ti}}
\end{equation*}%
Here, the notation indicates that $\Omega $ is a function of all of the
order parameters $\left\{ \psi _{ti}\right\} $. Let us suppose that the
system is described by some alternate set of order parameters, $\left\{ \phi
_{ti}\right\} $, which is complete in the sense that the two sets are
equivalent and the relation between them is invertible: $\phi _{ti}=\phi
_{ti}\left[ \psi \right] $ and $\psi _{ti}=\psi _{ti}\left[ \phi \right] $.
Then, it follows that%
\begin{eqnarray}
\frac{d\phi _{ti}}{dt} &=&-\Gamma \sum_{j}\frac{\partial \Omega \left[ \psi
_{t}\right] }{\partial \psi _{tj}}\frac{\partial \phi _{ti}}{\partial \psi
_{tj}}  \label{dyn1} \\
&=&-\Gamma \frac{\partial \Omega \left[ \phi _{t}\right] }{\partial \phi
_{tl}}\sum_{j}\frac{\partial \phi _{tl}}{\partial \psi _{tj}}\frac{\partial
\phi _{ti}}{\partial \psi _{tj}}  \notag
\end{eqnarray}

Suppose we did not know about this dynamics and simply wanted to write down
the steepest descent equations for these models. In that case, it is
necessary to specify a metric. We assume that the $\psi $-space is Euclidean
so that the distance between two sets of fields is%
\begin{eqnarray}
d^{2}\left[ \psi _{t}\left( \mathbf{r}\right) ,\psi _{t}^{\prime }\left( 
\mathbf{r}\right) \right]  &=&\int \left( \psi _{t}\left( \mathbf{r}\right)
-\psi _{t}^{\prime }\left( \mathbf{r}\right) \right) ^{2}d\mathbf{r} \\
&\rightarrow &\sum_{i}\left( \psi _{ti}-\psi _{ti}^{\prime }\right) ^{2} 
\notag
\end{eqnarray}%
This implies that%
\begin{equation}
d^{2}\left[ \phi _{t}\left( \mathbf{r}\right) ,\phi _{t}^{\prime }\left( 
\mathbf{r}\right) \right] \rightarrow \sum_{i}\left( \psi _{ti}\left[ \phi
_{t}\right] -\psi _{ti}\left[ \phi _{t}^{\prime }\right] \right) ^{2}
\end{equation}%
Assuming sufficient analyticity, this prescribes a Riemann geometry with
metric%
\begin{equation}
g_{lm}=\sum_{i}\frac{\partial \psi _{ti}\left[ \phi _{t}\right] }{\partial
\phi _{tl}}\frac{\partial \psi _{ti}\left[ \phi _{t}\right] }{\partial \phi
_{tm}}
\end{equation}%
In general, the steepest descent equations are 
\begin{equation}
\frac{d\phi _{ti}}{ds}=-\frac{g^{il}\frac{\partial \Omega \left[ \phi _{t}%
\right] }{\partial \phi _{tl}}}{\left( \frac{\partial \Omega \left[ \phi _{t}%
\right] }{\partial \phi _{ta}}g^{ab}\frac{\partial \Omega \left[ \phi _{t}%
\right] }{\partial \phi _{tb}}\right) }
\end{equation}%
Comparison of this to Eq.(\ref{dyn1}) shows that the dynamics is equivalent
to steepest descent with the relation between time and the distance
parameter being 
\begin{equation}
\Gamma dt=\frac{ds}{\left( \frac{\partial \Omega \left[ \phi _{t}\right] }{%
\partial \phi _{ta}}g^{ab}\frac{\partial \Omega \left[ \phi _{t}\right] }{%
\partial \phi _{tb}}\right) }.
\end{equation}

\bigskip

\bigskip

\bigskip

\bibliography{crystal_nucleation}

\end{document}